\documentclass[12pt,preprint]{aastex}

\slugcomment{Submitted to the Astrophysical Journal}

\lefthead{Wood et al.}
\righthead{Porosity of the Interstellar Medium}

\begin{document}

\title{Estimating the Porosity of the Interstellar Medium from 
Three-Dimensional Photoionization Modeling of H~{\sc ii} Regions}

\author {Kenneth Wood$^1$, L.M.~Haffner$^2$, R.J. Reynolds$^2$, 
John S. Mathis$^2$, and G. Madsen$^{2, 3}$ }

\altaffiltext{1}{School of Physics \& Astronomy, University of St. Andrews,
North Haugh, St Andrews, KY16 9SS, Scotland;
kw25@st-andrews.ac.uk}

\altaffiltext{2}{Astronomy Department, University of Wisconsin, 
475 N. Charter Street, Madison, WI 53706; haffner, reynolds, mathis, 
madsen@astro.wisc.edu}

\altaffiltext{2}{Anglo-Australian Observatory, PO Box 296, Epping, NSW 1710, 
Australia}

\email{kw25@st-andrews.ac.uk}

\begin{abstract}

We apply our three dimensional photoionization code to model Wisconsin 
H$\alpha$ Mapper observations of the H~{\sc ii} region surrounding 
the O9.5V star $\zeta$~Oph.  Our models investigate the porosity of the 
interstellar medium 
around $\zeta$~Oph and the effects of 3D densities on the H$\alpha$ 
surface brightness and variation in the [N~{\sc ii}]~$\lambda$6583/H$\alpha$ 
line ratio.   The 
$\zeta$~Oph H~{\sc ii} region has a well characterized ionizing source, 
so it is an excellent starting point for 3D models of diffuse ionized gas.  
We investigate various hierarchically clumped density structures, varying 
the overall smoothness within the clumping algorithm.  By simulating the 
observations, we can estimate the porosity of the medium in the vicinity of 
$\zeta$~Oph and 
find that within the context of our hierarchically clumped models, around 
50\% to 80\% of the volume is occupied by clumps surrounded 
by a low density smooth medium.  We also conclude that in order for O stars 
to ionize the diffuse Warm Ionized Medium, the O star environment must be 
more porous than that surrounding $\zeta$~Oph, with clumps 
occupying less than one half of the interstellar volume.  
Our clumpy models have irregular boundaries, similar to 
observed H~{\sc ii} regions. However, in observed H~{\sc ii} regions it 
is difficult to identify the precise location of the boundary 
because of the foreground and/or background emission from the 
widespread Warm Ionized Medium.  This complicates the interpretation of 
the predicted rapid rise of some emission line ratios near 
the edge of uniform density H~{\sc ii} regions and combined with the 
three dimensional clumpy nature of the interstellar medium may explain 
the apparent lack of distinctive emission line ratios 
near H~{\sc i} -- H~{\sc ii} interfaces.

\end{abstract}


\section{Introduction}

The interstellar medium (ISM) is observed to be clumpy on a very wide range 
of size scales from parsec scale molecular clouds to AU sizes 
inferred from interstellar scintillation measurements (Hill et al. 2004; 
Scalo \& Elmegreen 2005).  This clumpiness is 
observed to be self-similar over a range of scales with hierarchical 
or fractal dimensions in the range 2.2 to 2.7 
(Sanchez, Alfaro, \& Perez 2005).  
The complex structures are thought to 
result from mechanical and radiative energy 
input into the ISM from many processes including stellar winds, 
supernovae, shocks, 
photoionization, and cosmic rays.  This is reflected in the multi-phase 
nature of the 
ISM, where the hot ionized, warm ionized, warm neutral, cold neutral, and 
molecular gas co-exist.  There is considerable 
debate among theorists regarding the volume filling factors of each phase 
and the overall structure of the ISM.  Over the last decade large 
multiwavelength surveys of the ISM are providing new high spatial and 
spectral resolution observations against which observational signatures 
of theoretical models may be tested.  These velocity resolved surveys 
include neutral gas (Hartmann \& Burton 1997), molecular gas 
(Dame et al. 2001), and diffuse ionized gas (Haffner et al. 2003).  
The combination of the new datasets and advances in 3D dynamical and 
radiation transfer simulations now allows for critical testing of 
global models of the structure and dynamics of the ISM.  In this 
paper we use three-dimensional photoionization models to study the 
structure of a low density H~{\sc ii} 
region in the Galaxy on scales of tens of parsecs.  
This is a first step towards understanding 
the structure, clumpiness, and ionization of the large scale 
diffuse ionized gas in the ISM. 

Widespread diffuse ionized gas is a significant component of the ISM 
in the Milky Way (Kulkarni \& Heiles 1988; Reynolds 1995) and other 
galaxies (Rand 1997; Hoopes \& Walterbos 2003; Thilker et al. 2002; 
Collins et al. 2000).  The diffuse ionized gas, hereafter referred to 
as the warm ionized medium (WIM), in the Milky Way 
is revealed primarily through faint diffuse H$\alpha$ emission and 
pulsar dispersion measures. 
The physical characteristics of this gas, including its temperature 
structure and ionization state, as well as the spectrum of its 
ionizing sources are probed through analysis of optical forbidden 
emission lines.  
The H$\alpha$ data combined with the pulsar data 
suggest that the WIM has a volume filling 
factor of about 20\% (Reynolds 1991), and is one of the principal components 
of the ISM.  However, there is considerable uncertainty regarding the 
structure, dynamics, source of ionization, and heating of the WIM 
(e.g., see recent review by Reynolds, Haffner, \& Madsen 2002).  
The Wisconsin H$\alpha$ Mapper (WHAM) has completed an H$\alpha$ survey of 
the Northern sky (Haffner et al. 2003) revealing the distribution and 
kinematics of the WIM in the Galaxy for the first time.  
Several large regions are also being 
mapped with WHAM in other lines such as H$\beta$, He~{\sc i}~$\lambda$5876, 
[N~{\sc ii}]~$\lambda$6583, 
[S~{\sc ii}]~$\lambda$6716, [O~{\sc i}]~$\lambda$6300, and 
[O~{\sc iii}]~$\lambda$5007.  The data have 
revealed the clumpy and filamentary structure of the WIM on large 
scales, such as the bipolar loop structure in 
the Perseus Arm which extends $\pm 1$~kpc 
above and below the Galactic plane (Haffner at al. 1999; Madsen 2004).  
In addition to the large scale structures, WHAM has mapped several 
nearby low density H~{\sc ii} regions with sizes of tens of parsecs 
(e.g., Haffner et al. 1999).  
Three-dimensional photoionization modeling of these 
H~{\sc ii} regions may allow us to estimate the structure and porosity 
of the ISM on these scales.  Such models of well-defined H~{\sc ii} 
regions with known ionizing sources are a first step in modeling the 
WIM on larger scales with multiple and unidentified ionizing sources.  

In this paper we study the three-dimensional structure of the 
H~{\sc ii} region surrounding the O9.5V star $\zeta$~Oph, 
incorporating recent WHAM observations in the lines of H$\alpha$, 
[N~{\sc ii}]~$\lambda$6583, and [S~{\sc ii}]~$\lambda$6716 
(Haffner et al. 1999; Baker et al. 2005).  We use the 
three-dimensional photoionization code of Wood, Mathis, \& Ercolano (2004) 
to model the observed H$\alpha$ intensity and line ratios in order to 
estimate the clumpiness of the ISM around $\zeta$~Oph.  Our models 
enable us to investigate the penetration and escape of ionizing 
radiation from clumpy H~{\sc ii} regions into the more widespread WIM.  
In addition, we study the observational signatures of ionized/neutral 
interfaces in smooth and clumpy models.  
Section~2 presents the H$\alpha$ intensity 
and line ratio maps of the $\zeta$~Oph region, \S3 summarizes model 
predictions for line ratios in H~{\sc ii} regions and at ionized/neutral 
interfaces, \S4 and \S5 present smooth and 
clumpy models for the observations, and we summarize our findings in \S6.

\section{The $\zeta$~Oph H~{\sc ii} Region}

Figure~1 shows the structure of the H~{\sc ii} region around $\zeta$~Oph.  
The upper left panel shows the H$\alpha$ intensity from the SHASSA 
survey (Gaustad et al. 2001; Finkbeiner 2003) 
with an angular resolution of $0.8\arcmin$. 
The upper right panel shows the same region at $1^\circ$ resolution 
from the WHAM Northern Sky Survey (Haffner et al. 2003). We have 
interpolated the irregularly-spaced WHAM observations onto a 
regular $0\fdg5 \times 0\fdg5$ grid.  
The location of $\zeta$~Oph is shown with a cross.  
The WHAM image shows that the overall structure of the H$\alpha$-emitting 
gas is remarkably circularly symmetric, although there are 
strong fluctuations on smaller angular scales as shown in the 
SHASSA image.  Diffuse H$\alpha$ emission associated with the 
$\zeta$~Oph H~{\sc ii} region is traced out to around 
$6^\circ$ from the star, corresponding to a radius of about 15~pc, before 
trailing off into more diffuse foreground/background emission 
toward higher Galactic longitudes (on the left).  
On the other side of the H~{\sc ii} region the H$\alpha$ 
intensity becomes confused with that from the $\delta$~Sco 
H~{\sc ii} region (see Haffner et al. 2003).  
We do not consider data from this region in our analysis. 

The sightline towards $\zeta$~Oph is well studied with many 
absorption line determinations of interstellar abundances 
(e.g., Morton 1975; Cardelli et al. 1993, 1994; Howk \& Savage 1999).  
The star itself is catalogued as O9.5V, with an estimated 
temperature in the range $32\, 000 < T_\star < 33\, 000$~K 
(Code et al. 1976; Markova et al. 2004), 
and a distance of 140~pc (Perryman et al. 1997; Howk \& Savage 1999).  
$\zeta$~Oph has a heliocentric radial velocity of $-10.7$~km~s$^{-1}$, which 
is $+3.3$~km~s$^{-1}$ relative to the local standard of rest.  
The WHAM data shows that the ionized gas in the H~{\sc ii} region 
lies at 0~km~s$^{-1}$. Its proper 
motion from {\it HIPPARCOS} is $13\pm 1$~mas~yr$^{-1}$ in right ascension 
and $25.5\pm 0.5$~mas~yr$^{-1}$ in declination, resulting in a space 
velocity of $19.3\pm 0.4$~km~s$^{-1}$ relative to the LSR for a distance 
of 140~pc.   With such a velocity, $\zeta$~Oph travels more than 19~pc 
in $10^6$~years, larger than the 15~pc radius of its 
own H~{\sc ii} region.  Therefore over its lifetime, 
it appears that $\zeta$~Oph has moved 
from its birthplace and is now ionizing a different region of the ISM from 
where it was born.  We will return to this point when discussing our 
photoionization models below.

The H$\alpha$ maps have not been corrected for foreground extinction by 
interstellar dust that may lie between us and the H~{\sc ii} region.  
In particular, the dust lane cutting across the lower left quadrant in 
the SHASSA image is likely in  the foreground and not and integral part 
of the H~{\sc ii} region.  
Future H$\beta$ observations with WHAM will allow for 
dust correction and the determination of which features are internal to 
the H~{\sc ii} region and which are due to foreground dust.  
However, the [N~{\sc ii}]/H$\alpha$ and [S~{\sc ii}]/H$\alpha$ 
line ratio maps will not be 
affected by dust since H$\alpha$, [N~{\sc ii}], and [S~{\sc ii}] 
are very close in wavelength.  

In addition to the H$\alpha$ maps, Fig.~1 shows WHAM line ratio maps of 
[N~{\sc ii}]/H$\alpha$ and [S~{\sc ii}]/H$\alpha$.  
These maps show these line ratios increase with distance 
away from $\zeta$~Oph.  Figure~2 shows 
a scatter plot of H$\alpha$, [N~{\sc ii}]/H$\alpha$ and 
[S~{\sc ii}]/H$\alpha$ against radial distance from $\zeta$~Oph.  The line 
ratios increase fairly smoothly with increasing distance and do not show 
the rapid increase expected towards the edge of a uniform density 
H~{\sc ii} region (see \S3).  
For comparison with models, in the rest of the paper when showing the 
WHAM data we show the mean radial surface brightness and line ratios 
and the dispersions about the mean.  The mean radial surface brightness 
is the average value for the H$\alpha$ intensity (or line ratios) 
in circular annuli centered on $\zeta$~Oph.

\section{Interface Emission: Predictions and Observations}

One dimensional photoionization models predict rapid increases in the 
ratios of the projected intensities of [O~{\sc i}], [N~{\sc ii}], and 
[S~{\sc ii}] relative to H$\alpha$ towards the edge of uniform density 
H~{\sc ii} regions (e.g., Henney et al. 2005).  
The increased line ratios occur in the transition 
region at the edge of the Str{\" o}mgren sphere where the fraction of 
neutral gas is increasing 
and the temperature is rising due to hardening of the radiation 
field.  The wavelength dependence of H$^0$ opacity allows the highest 
energy photons to penetrate the largest distances, resulting in a 
rising temperature in the interface region (e.g., Osterbrock 1989).  

Observations of the diffuse ISM near the Galactic plane 
show low [O~{\sc i}]/H$\alpha$ line ratios (Reynolds et al. 1998), 
suggesting that interfaces between ionized and neutral gas have either 
not been detected or do not contribute significantly to the emission.  
Photoionization models can 
explain the low [O~{\sc i}]/H$\alpha$ observations if the WIM is 
highly ionized (${\rm H}^0/{\rm H} < 0.1$ throughout), implying 
that ionizing radiation escapes from fully 
ionized (i.e., density bounded) H~{\sc ii} regions having few 
H~{\sc i} -- H~{\sc ii} interfaces 
(e.g., Mathis 1986; Domgoergen \& Mathis 1994; 
Mathis 2000; Sembach et al. 2000).  
Also, recent work by Giammanco et al. (2004) 
showed that if a clumpy H~{\sc ii} region is comprised of a mixture of 
fully and partially ionized 
spherical clouds, the line ratios increase more slowly with radius 
than in uniform density, ionization bounded models (e.g., see the radial 
dependence of [O~{\sc i}]/H$\beta$ 
in their Fig.~6).  This is because the combination of many spherical clouds 
along a given sightline through the H~{\sc ii} region sample a wide 
range of ionization parameters and hence the line ratios do not show the 
rapid increase of uniform, ionization bounded models.  

While the predicted rapid rise of temperatures and line ratios 
at the edges of H~{\sc ii} regions has not been observed (e.g., Pauls \& 
Wilson 1977), gradual rises 
in the line ratios have been observed with height above the midplane 
ionizing sources in the Milky Way (Haffner et al. 1999) and several 
other edge-on galaxies (Rand 1998; Otte et al. 2001, 2002).  The elevated 
[N~{\sc ii}]/H$\alpha$ line ratios observed at large distances from 
the midplane in the Perseus Arm (Haffner et al. 1999) have been 
interpreted as due to an extra heat source in addition to 
photoionization heating (Reynolds et al. 1999).  This additional heat 
source must have the property that it dominates over photoionization 
heating at low densities.  A constant energy source or one that is 
proportional to $n_{\rm e}$ would suffice, since photoionization heating 
is proportional to $n_{\rm e}^2$.  Candidates for the additional 
heating include photoelectric heating from grains (Reynolds \& Cox 1992), 
magnetic reconnection (Raymond 1992), dissipation of turbulence 
(Minter \& Spangler 1997), shocks and cooling hot gas (Slavin, Shull, 
\& Begelman 1993; Collins \& Rand 2001).  
However, hardening of the radiation field also increases the 
temperature of photoionized gas and may help to explain in part 
some of the observed 
elevated line ratios (e.g., Bland-Hawthorn, Freeman, \& Quinn 1997; 
Elwert 2003; Wood \& Mathis 2004). 

In this paper we present models for the H~{\sc ii} region associated 
with $\zeta$~Oph.  Additional heating will not be important for this 
source since the gas density is larger and the 
line ratios are much lower than observed at high 
latitude in the Perseus arm.  
The WHAM observations presented in Figures~1 and 2 
show that [N~{\sc ii}]/H$\alpha$ increases away from 
the source, but does not appear to show the rapid rise at the edge of the 
H~{\sc ii} region predicted by spherical models with uniform density 
(see \S4.1 below).  
To understand this behavior we have investigated photoionization models 
that incorporate 
various smooth and clumpy density structures for the H~{\sc ii} region and 
the effects of diffuse foreground/background emission.   
We discuss the plausibility of the smooth models and the insight 
three-dimensional models can provide into the structure of the ISM 
around $\zeta$~Oph and the escape of ionizing photons from 
H~{\sc ii} regions. Our 3D models produce similar results to those 
of Giammanco et al. (2004) for radial line ratio gradients, but our 3D 
radiation transfer allows for shadowing of clumps and multiple clumps 
along any sightline from the star.

\section{Photoionization Models}

To model the WHAM observations of the $\zeta$~Oph H~{\sc ii} region, 
we use the three-dimensional photoionization code described in Wood, 
Mathis, \& Ercolano (2004).  
This code calculates the 3D ionization and temperature 
structure for arbitrary geometries and illuminations, keeping track of 
the ionization structure of H, He, C, N, O, Ne, and S.  
The opacity is from to H$^0$ and He$^0$ and we only consider photon 
energies in the range 13.6~eV to 54~eV.  Heating is from photoionization 
of H and He and cooling is from H and He recombination, free-free emission, 
and collisionally excited line emission from C, N, O, Ne, and S.  The 
output of our code is the ionization and temperature structure from which 
we calculate emissivities and intensity maps for the various emission lines 
we wish to study.  A complete description of the code is presented in 
Wood et al. (2004).  

We perform the 
radiation transfer for stellar and diffuse recombination 
ionizing radiation on a 65$^3$ linear Cartesian grid.  Due to the 
resolution of our grid, we do not consider emission 
from cells that are more neutral than ${\rm H}^0/{\rm H}\ge 0.25$ 
(see discussion in Wood et al. 2004).  
These cells lie towards the edge of the ionized volume and 
account for less than 20\% of the ionized cells in our ionization bounded 
models.  We have constructed intensity and line ratio maps including 
these cells and find that our results for the [N~{\sc ii}]/H$\alpha$ 
line ratio maps do not change at the $\sim 5$\% level.  Moreover, 
in the cells at the ionization boundary our code is incomplete, 
since we ignore the effects 
of dynamics and shocks at the ionization front (Henney et al. 2005).  Higher 
resolution simulations ($128^3$ grids) do not appreciably change our results, 
again at around the $\sim5$\% level.

Our code performs well compared to other photoionization codes in 
predicting ionization fractions, temperatures, and line 
strengths.  Here we focus on modeling the observed 
H$\alpha$ intensity and [N~{\sc ii}]/H$\alpha$ maps and do not consider 
the [S~{\sc ii}]/H$\alpha$ observations.  Modeling emission from sulfur 
is problematic due to the unknown dielectronic recombination 
rates for S (Ali et al. 1990).  In addition we do not consider the 
effects of dust within the H~{\sc ii} region and note that our 
temperatures may be slightly higher than calculations that include 
cooling from more elements (e.g., see Sembach et al. 2000).  
Since the goal is to investigate the 3D structure of 
the H~{\sc ii} region by modeling variations in the line 
ratios towards the edge of the H~{\sc ii} region, ignoring these effects 
will not change our conclusions on the H~{\sc ii} region density 
structure.  Inclusion of these additional effects 
will be required for more comprehensive models of future observations 
of H$\beta$, [O~{\sc i}], [O~{\sc ii}], [O~{\sc iii}], and He~{\sc i}.  
In particular, we may 
be able to model data from several H~{\sc ii} regions to place 
constraints on the S dielectronic recombination rates. 

Relative to H, the adopted abundances for He, C, N, O, Ne, and S in our 
models are 0.1, 140~ppm, 75~ppm, 319~ppm, 117~ppm, and 18~ppm.  These 
are interstellar gas phase abundances from the compilation of Sembach et al. 
(2000).  
The ionizing spectrum is taken to be that of a 32~000~K WM-basic model 
atmosphere from the library of Sternberg, Hoffmann, \& Pauldrach (2003).  
This temperature 
is within the range estimated for an O9.5V star (see compilation of 
temperature/spectral type scales and discussion 
in Harries, Hilditch, \& Howarth 2003, Table~4).  The 
WM-basic model atmospheres include the effects of line blanketing and 
stellar winds (Pauldrach et al. 2001).  
The ionizing luminosity and density are varied in the models to 
reproduce the observations of H$\alpha$ and [N~{\sc ii}]/H$\alpha$.  
The following sections present results for smooth 
models (with and without density gradients) and three-dimensional, 
hierarchically clumped models.  

\subsection{Smooth Spherically Symmetric Models}

Our spherically symmetric models are centered on a source with an 
ionizing photon luminosity $Q$~(s$^{-1}$), 
and the density within the cloud (cm$^{-3}$) is given by 
\begin{equation}
n(r) = n_0 \left( {{r}\over {1\,{\rm pc}}} \right)^{-p}\; .
\end{equation}
where the parameter $p$ controls the radial density gradient 
within the cloud.  
In all models (smooth and clumpy) we have left the inner 10\% of the 
grid empty, which prevents the clumping algorithm randomly placing 
the star in the center of a dense clump.  The grid is 60~pc on a side and 
contains $65^3$ cells, while the observed $\zeta$~Oph H~{\sc ii} region 
is about 
40~pc in diameter but has some extensions to greater distances.
We have investigated many ($p$, $n_0$, $Q$) combinations for ionization and 
density bounded models and show those that reproduce the overall 
H$\alpha$ intensity level observed by WHAM.  Table~1 summarizes 
the range of adopted model parameters and the results of the models 
are compared to the observations in figure~3.

The radial variation of H$\alpha$ and [N~{\sc ii}]/H$\alpha$ 
may be influenced by the presence of foreground and background 
emission.  This is illustrated in Figure~3 which shows 
models without (left panels) and with (right panels) the 
addition of a constant intensity representing foreground/background 
emission.  Both the smooth (shown here) and clumpy models (shown in \S4.2) 
have been convolved 
with a Gaussian beam to simulate the $1^\circ$ angular resolution of WHAM.  
The diffuse background is taken to be 1.5~R at H$\alpha$ and 
0.6~R for the [N~{\sc ii}] simulation.  These values provide a good fit 
to the observations (see Fig.~2) 
and also are typical for the diffuse interstellar 
medium near the Galactic latitude ($b=24^\circ$) of $\zeta$~Oph 
(see also Haffner et al. 2003; Madsen 2004).  
The addition of diffuse foreground/background emission 
(which is an unavoidable contaminant in real observations) 
clearly suppresses the very large line ratios that are predicted in the 
faintest outermost regions ($\sim 20$~pc) of the H~{\sc ii} region.  
However, even with the addition of foreground/background 
emission, the uniform density, ionization bounded model 
(i.e., a classic Str{\" o}mgren sphere) exhibits a rise in the 
[N~{\sc ii}]/H$\alpha$ at the edge of the H~{\sc ii} region that is 
steeper than the observations.  The H$\alpha$ intensity gradient 
in such models is also much steeper at the edge of the ionized region 
than observed (see 
also the giant H~{\sc ii} region models of Giammanco et al. 2004).  
Density bounded or ``leaky H~{\sc ii} 
region'' models (Fig.~3) do not exhibit the rapid rise in line ratios 
because there is no ionized/neutral interface in the simulations 
(e.g., Mathis 2000; Sembach et al. 2000).  
The density bounded models shown in Fig.~3 
allow about 40\% of the ionizing photons to escape.  
Changing the radial density gradient from uniform ($p=0$) 
to steeper laws ($p=1$, 2) changes the H$\alpha$ radial surface 
brightness profile.  Models with gradients steeper than $p=1$ do not 
match the WHAM data.  Models with $p=1$ appear to give a reasonable match 
to the H$\alpha$ intensity, but all the density bounded models 
underpredict the [N~{\sc ii}]/H$\alpha$ observations.

As mentioned previously, $\zeta$~Oph is at 
a different radial velocity from its associated H~{\sc ii} region, 
so models that place the star at the centre of a spherical cloud with 
radial density gradients seem contrived.  A more realistic 
scenario is that $\zeta$~Oph is ionizing a clumpy medium and we explore 
3D cloud geometries and the effects on the intensity and line ratio maps 
in the next section.

\subsection{Hierarchically Clumped Models}

There is much evidence for hierarchical structure in the ISM, 
with surveys of clouds revealing fractal structures 
(e.g., Elmegreen \& Falgarone 1996).  
We therefore investigate 3D H~{\sc ii} regions which have hierarchically 
clumped density structures.  
A 3D hierarchical density structure is created using the algorithm 
presented by Elmegreen (1997).  We use five hierarchical levels and 
the density in the grid is set as follows.  At the first level we randomly 
cast $N$ points with $x$, $y$, and $z$ coordinates in the range (0,1).  
At each subsequent hierarchical level we cast $N$ random points around each 
of the points cast at the previous level.  The casting length for the 
$x$, $y$, and $z$ points at each subsequent level 
is in the range $\pm \Delta^{(1-H)} /2$, where $H$ is the hierarchical level 
being cast, and $f=\log N / \log \Delta$ is the 
fractal dimension of the hierarchical structure.  The casting length, 
$\Delta$, gets smaller for subsequent hierarchical levels.  Note that in 
this description $\Delta$ is dimensionless and we construct the hierarchical 
grid within a cube with side of unit length.  Actual dimensions are 
constructed by mapping the hierarchical grid onto our 3D density grid, so 
that the unit cube corresponds to the physical size of our density grid.  
We use a fractal dimension $f=2.6$, appropriate for interstellar clouds 
(e.g., Sanchez, Alfaro, \& Perez 2005).  Using this algorithm, 
the density of a cell in our grid is proportional to the number of points 
cast at the final level that lie in the cell.  If points are cast beyond 
the $x$, $y$, or $z$ boundaries of the grid, they are added into the 
corresponding cells on the opposite side of the grid, e.g. if a point is 
cast at $x=x_{\rm max}+\epsilon$, we change its coordinate to be 
$x=\epsilon$ and similarly for $y$ and $z$ coordinates.  
Figure~4 illustrates this algorithm 
in 2D for a three tier hierarchical scheme, with four random points 
cast at each level.  The filled in squares are the first castings, 
diamonds the second level, and crosses at the third level.  The dotted 
lines show overlaid grid cells.  We assume that a fraction, 
$f_{\rm smooth}$, of the average density is present in every cell.  The 
remaining mass is distributed in proportion to the number of crosses that 
are in each cell.

The smooth component represents unresolved 
fractal structure that cannot be resolved within our grid.  In 
real nebulae the hot, shocked stellar winds may create very low density 
regions between the clumps of gas.  
The resulting clumpy density grid is renormalized so that it has 
the same mean density structure as the smooth grid. We generally 
adopt $f_{\rm smooth}=1/3$, but also consider the case where there is 
zero smooth density component.  Our adopted value of $f_{\rm smooth}$ 
comes from radiation transfer models of the penetration of radiation 
into turbulent cloud models from hydrodynamic simulations.  A value 
of $f_{\rm smooth}=1/3$ provides a good match for the internal 
intensity levels between the hydrodynamical simulations and density 
structures generated with the fractal algorithm (Bethell et al. 2004).  

To investigate the 
clumpiness of the ISM around $\zeta$~Oph we varied $N$ at the first 
hierarchical level, using $N_1=32$, 64, 128, 256, and 512,  but keeping 
$N=32$ at all four subsequent hierarchical levels.  With this modification 
to the clumping algorithm the relation between casting length, $\Delta$, 
and fractal dimension, $f=\log N / \log\Delta$, is determined assuming 
$N=32$.  Increasing $N_1$ at the 
first level gives a progressively smoother density structure at large 
spatial scales, but maintains the observed fractal dimension of 
interstellar clouds on smaller scales.  We have not investigated the 
effects on the ionization structure and line ratios of changing the value 
$N=32$ at the second and higher hierarchical levels.  Increasing $N$ will 
give a larger dynamic range and finer structure in the hierarchical density 
grid, because the total number of points used to determine the density is 
given by the number of castings at the final level, $N_1\times N^4$.  
The choice of $N=32$ is not observationally motivated, but is determined 
by the resolution of our density grid.  Using very large values of $N$ is 
not justified because the fine structure will not be resolved by our grid.  

The algorithm produces maximum density contrasts 
between the densest clumps and the 
smooth component of 65, 35, 20, 10, for $N_1 = 32$, 64, 128, and 512 
respectively.  A small $N_1$ allows ionizing radiation to penetrate 
further between the clumps than in a uniform medium (Elmegreen 1997).  
The fraction of cells that contain only the smooth component of density, 
without any of the points that were cast randomly, is 
65\%, 45\%, 20\%, and 0.1\% for these models.  
If the density in the smooth component is almost zero, ionizing 
radiation can escape 
along the empty sightlines into the larger scale ISM.  

Having a smooth component of zero density 
did not significantly change the morphology 
of our H$\alpha$ maps compared to the $f_{\rm smooth}=1/3$ that we adopt 
because the emission is dominated 
by the higher density ionized clumps.  For $N_1=32$ and 
$f_{\rm smooth}=1/3$, the smooth component 
accounts for about 1/3 of the H$\alpha$ intensity level.  The 
smooth contribution 
to the H$\alpha$ intensity decreases with increasing $N_1$.  
Elmegreen (1997) speculates that much of the volume of a fractal 
ISM is at very low density, like our models with no smooth component.

We investigated 
many different clumpy models for each $N_1$ value by changing 
the random number seed used for setting up the hierarchical density 
grid.  For a given $N_1$, each model has a different density 
structure and hence slightly different H$\alpha$ and [N~{\sc ii}] 
intensity profile due to the random casting of clumps around the star, 
but the overall features in the intensity maps are similar.  
For each $N_1$ of our clumpy models, 
Figures~5 through 8 show one pixel thick 
slices through one of the density grids and slices 
showing the hydrogen ionization fraction ${\rm H}^0/{\rm H}$.  
The figures also show the H$\alpha$ intensity and [N~{\sc ii}]/H$\alpha$ 
line ratio maps, {\it without} the addition of uniform foreground/background 
emission.  
Figures 9 to 12 show the radial variation of H$\alpha$ and 
[N~{\sc ii}]/H$\alpha$ {\it with} 
the addition of uniform foreground/background 
emission.  For each $N_1$ model, the figures show one model with the 
standard deviations about the mean and also the radial profiles for 
simulations of five different density grids.

\section{Results}

Our models are constrained by the emission line intensities observed at 
various angles from the star (or distances projected onto the sky) and their 
dispersions.  For all clumpy models the ionizing luminosity is 
$Q=8\times 10^{47}\,{\rm s}^{-1}$ and the mean density 
of the H~{\sc ii} region is $n=2\,{\rm cm}^{-3}$.  Both the luminosity and 
density are about 50\% smaller than estimated by Elmergreen (1975) from 
analysis of older H$\alpha$ data from Reynolds et al. (1974).  
Our estimates for the luminosity and density are 
well constrained by the intensity and angular extent of the H$\alpha$ data.
We find that $N_1\sim64$, followed by coarser clumping on 
the smaller scales, provides the best fit to the WHAM data.  This implies 
that the real ISM in this direction is not hierarchical, but is smoother 
on a scale of $\sim 10$~pc than at smaller scales.  This result reflects the 
general circular symmetry in the WHAM H$\alpha$ image in Fig.~1, but the 
small scale irregularities are not revealed in the WHAM data because of 
its relatively coarse angular resolution ($1^\circ$).

As expected, Figures 5 through 12 show that 
the clumpy model that most resembles a Str{\" o}mgren sphere has 
$N_1=512$, since distributing many random points approximates a 
uniform distribution.  
This model has a somewhat irregular boundary, but there 
are no sightlines from the star that do not intersect dense clumps, 
so overall the model has a very smooth structure and exhibits the 
large increase in the [N~{\sc ii}]/H$\alpha$ line ratio at the edge of 
the H~{\sc ii} region.  As $N_1$ is reduced, the medium becomes less smooth 
and there are sightlines from the star through which ionizing radiation 
can traverse large distances before intersecting a clump, or not 
intersect any clumps.  Models with $N_1=32$ show a gradual 
radial decline of the H$\alpha$ intensity.  
As $N_1$ is increased, the fraction of low density sightlines 
decreases and the H$\alpha$ exhibits a steeper radial gradient.
The concavity of the radial H$\alpha$ surface brightness profiles 
change for larger $N_1$ and approach that of the uniform density 
Str{\" o}mgren sphere in Figure~3.  The trends in our models are 
similar to the radial surface brightness profiles and line ratios 
found by Giammanco et al. (2004) in their H~{\sc ii} region models 
comprising spherical blobs surrounding an ionizing source.

Within the context of our clumping algorithm, we find that models with 
$N_1=32$ produce too much H$\alpha$ at large radii and variations about  
the mean radial intensity are larger than observed.
Models with $N_1\ge 256$ resemble Str{\" o}mgren spheres, 
producing too steep a gradient of the H$\alpha$ and a rapid rise of 
[N~{\sc ii}]/H$\alpha$ at the edge of the 
H~{\sc ii} region and small variations about the mean values.  Models with 
$N_1=64$ and 128 provide better matches to the WHAM data in reproducing 
the observed levels of H$\alpha$ and [N~{\sc ii}]/H$\alpha$, their radial 
variations, and standard deviations.

While the intensity maps for 
this low density H~{\sc ii} region model are not very sensitive to the 
smooth component, the smooth component has a large influence 
on the ionizing radiation that may escape the H~{\sc ii} region and 
ionize the larger scale diffuse ionized gas.  Models with $N_1=32$, 
64, and 128 and zero smooth density component ($f_{\rm smooth}=0$) 
typically allow around 
10\%--25\%, 3\%--15\%, and 1\%--3\% 
respectively of the stellar ionizing photons to 
escape the H~{\sc ii} region directly along the zero density paths from 
the star.  The large variations in the escape fractions for each $N_1$ 
value are due to the random placing of clumps around the star.  
Our 3D photoionization models therefore confirm the analysis 
of Elmegreen (1997) that a fractal ISM will allow for ionizing photons 
to traverse much larger distances than a smooth density structure.  
Further clearing of a fractal ISM by feedback from photoionization 
(Dale et al. 2005) and stellar winds will likely provide 
an even larger escape fraction from H~{\sc ii} regions than those 
determined from our models.

Since the widespread H$\alpha$ emission from the Galactic WIM 
requires about 15\% of the ionizing photons from Galactic O stars 
(Reynolds 1995), it 
appears that a 3D fractal ISM structure could allow for such 
leakage from H~{\sc ii} regions.  Our hierarchical density 
models for the ISM around $\zeta$~Oph with 
$N_1=64$ and 128 typically provide escape fractions of less than 15\%, and 
even lower if the 
density in the smooth component is not zero.  Therefore, the 
ionizing sources responsible for powering the WIM in 
the Galaxy must reside in regions of the Galaxy that are more 
porous than the region around $\zeta$~Oph.  Perhaps 
their environment has been shaped more by turbulent motions or are 
regions cleared out by stellar winds, thus enabling the escape of 
ionizing photons.  Future models of WHAM 
observations of other H~{\sc ii} regions will enable us to determine 
whether the porosity levels for the ISM around $\zeta$~Oph are typical.

Our models placed the star at the 
center of a fractal cloud, so models where the star is at the edge of 
the cloud will allow for larger leakage, since the cloud no longer 
covers the whole sky as seen from the star.   
The major question from this work 
relating to ionization of the diffuse ionized gas 
is {\it what is the density of the smooth component in fractal models?}  
We cannot determine how much H$\alpha$ emission from the WIM comes 
from an extensive, relatively smooth component and how 
much is from ionized clumps and the ionized faces of clouds.  Further 
progress to answer this question may be made by conducting 
photoionization simulations on the 3D density grids being produced by 
global models of the ISM (e.g., de~Avillez \& Berry 2001).  Such models 
predict the ISM density structure and the locations of ionizing 
sources and their validity may be tested against observations through 
3D photoionization simulations.

\section{Summary}

We have presented smooth and clumpy 3D photoionization 
models for the $\zeta$~Oph 
H~{\sc ii} region.  We find that a star with $T=32\,000$~K and ionizing photon 
luminosity $Q=8\times 10^{47}\,{\rm s}^{-1}$ and mean density 
of the H~{\sc ii} region of $n\approx 2\,{\rm cm}^{-3}$ reproduce 
the observations.  These numbers are 
well constrained by the WHAM data.  The stellar temperature and ionizing 
luminosity are remarkably close to recent determinations of stellar 
properties for O9.5V stars (Martins et al. 2005, Table~4).  
Larger densities require a larger 
ionizing luminosity to match the extent of the H~{\sc ii} region, but then 
the H$\alpha$ values are larger than observed.  Smaller densities have 
the opposite effect, requiring a lower ionizing luminosity to match the 
extent of the H~{\sc ii} region, but then the simulated H$\alpha$ intensity 
is much smaller than observed.
 
Smooth models that best match the observations are 
density bounded and have a radial density gradient that falls off no 
faster than $r^{-1}$.  However, the velocity offset between $\zeta$~Oph 
and its H~{\sc ii} region suggest that these uniform models 
are unrealistic as they 
require the chance alignment of the star at the center of the cloud.  
The inclusion of foreground/background emission suppresses the rapid rise 
in the [N~{\sc ii}]/H$\alpha$ line ratio predicted at the edge of uniform 
Str{\" o}mgren spheres.

Hierarchically clumped models, which reproduce the observed fractal 
structure in the ISM, 
also reproduce the H$\alpha$ surface brightness and the 
shallow gradient  of [N~{\sc ii}]/H$\alpha$ away from $\zeta$~Oph.  
The best models suggest that around 20\% ($N_1=128$) to 50\% ($N_1=64$) 
of the H~{\sc ii} region volume is occupied by gas distributed in 
a smooth component or at very low density, with the 
remainder in clumps.  Three-dimensional 
models that are more uniform do not match the observations.  

The large, low density regions in 3D models 
could allow ionizing photons to 
escape at levels that sustain the ionization of the 
large scale diffuse ionized gas in the Galaxy.  However, the 
escape fractions from our preferred clumpy models for $\zeta$~Oph 
are less than the 15\% escape fraction required to ionize the WIM.  
Therefore we conclude that the stars responsible for ionizing the WIM 
must reside in regions of the ISM that are more porous than 
that surrounding $\zeta$~Oph, possibly evacuated by supernovae and/or 
stellar winds.  A more detailed investigation into the leakage of 
ionizing photons to the WIM requires consideration of dynamical clearing 
and the time evolution of the ionizing luminosity of O stars.

Future work will include more comprehensive modeling of the planned 
mapping observations of $\zeta$~Oph in additional emission lines.  
The WHAM survey has mapped out 
several other low emission measure H~{\sc ii} regions 
and 3D modeling will allow us to test whether the porosity values we 
estimate for the $\zeta$~Oph region are common in other regions.  
Modeling the [S~{\sc ii}] emission from many such regions may help to 
place constraints on the as yet unknown dielectronic recombination 
rates for sulfur.  From a theoretical perspective, 
global dynamical models of the ISM can be tested by applying 3D 
photoionization codes and comparing their observational signatures 
with the high resolution observations now available of 
diffuse ionized gas in our own and other galaxies. 

\acknowledgements
We thank Lynn Matthews and Jane Greaves 
for comments on an early version of this paper.  
We acknowledge funding from a PPARC Advanced Fellowship (KW) and a 
PPARC Visiting Fellowship to the University of St Andrews (JSM, RJR, GM).  
WHAM is funded by grants from the NSF (AST~0204973) 
and the University of Wisconsin Graduate School.

\pagebreak

\begin{table}
 \caption{Smooth Model Parameters}
 \label{smooth_models}
 \begin{tabular}{@{}lcccccc}
  \hline
   $n_0$ & $p$ & $R_{\rm max}$ 
   & $Q$ & Ionization\\
   $({\rm cm}^{-3})$ &  &  (pc) 
   & $(10^{47}\,{\rm s}^{-1})$ & bounded?\\
  \hline
  2.0 & 0 & 30 & 8 & yes\\
  2.0 & 0 & 18 & 20 & no\\
  7.5 & 1 & 18 & 20 & no\\
  98.0 & 2 & 18 & 20 & no\\
  \hline
 \end{tabular}
\end{table}

\pagebreak

\begin{figure}
\plotone{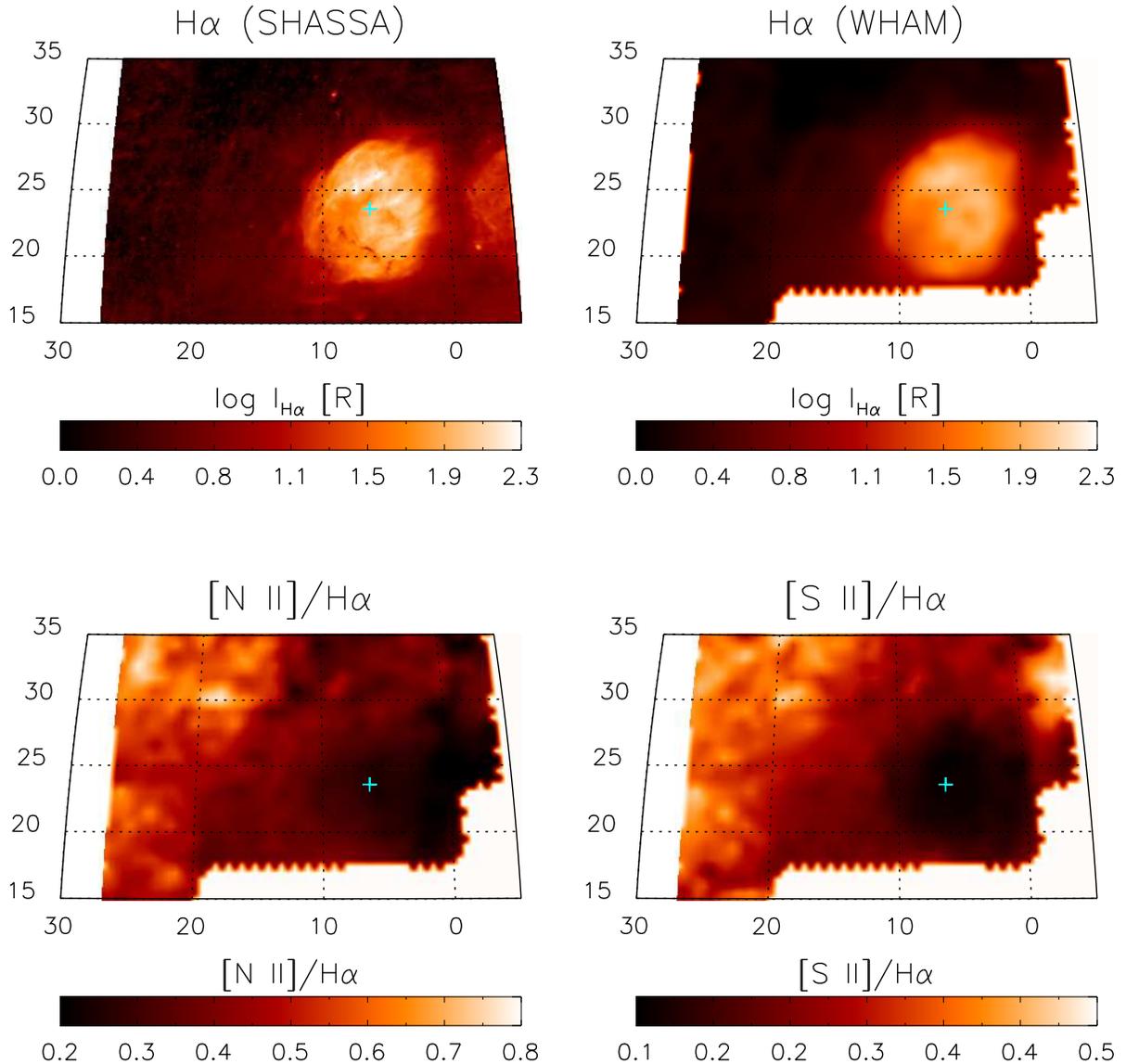}
\caption{Views of the ionized gas near $\zeta$ Oph (Sharpless 27). The top 
panels display the logarithm of the H$\alpha$ emission from 
(\emph{left}) a moderate-resolution ($\sim 8\arcmin$) image provided by 
SHASSA (Gaustad et al. 2001; Finkbeiner 2003) and (\emph{right}) a 
lower-resolution representation from WHAM.  The WHAM spectra have been 
integrated between $-25$ km s$^{-1} < v_\mathrm{LSR} < +25$ km s$^{-1}$ 
for these images. The lower panels show the (\emph{left}) [N 
II]/H$\alpha$ and (\emph{right}) [S II]/H$\alpha$ ratios of the region. 
The location of $\zeta$ Oph is denoted by the '+' in each panel.  
Part of the $\delta$~Sco H~{\sc ii} region may be seen at the right 
edge of the SHASSA image. All four images are displayed with an 
Aitoff-Hammer projection.  The axis labels are Galactic co-ordinates 
(latitude vs. longitude) in units of degrees.}
\end{figure}

\begin{figure}
\plotone{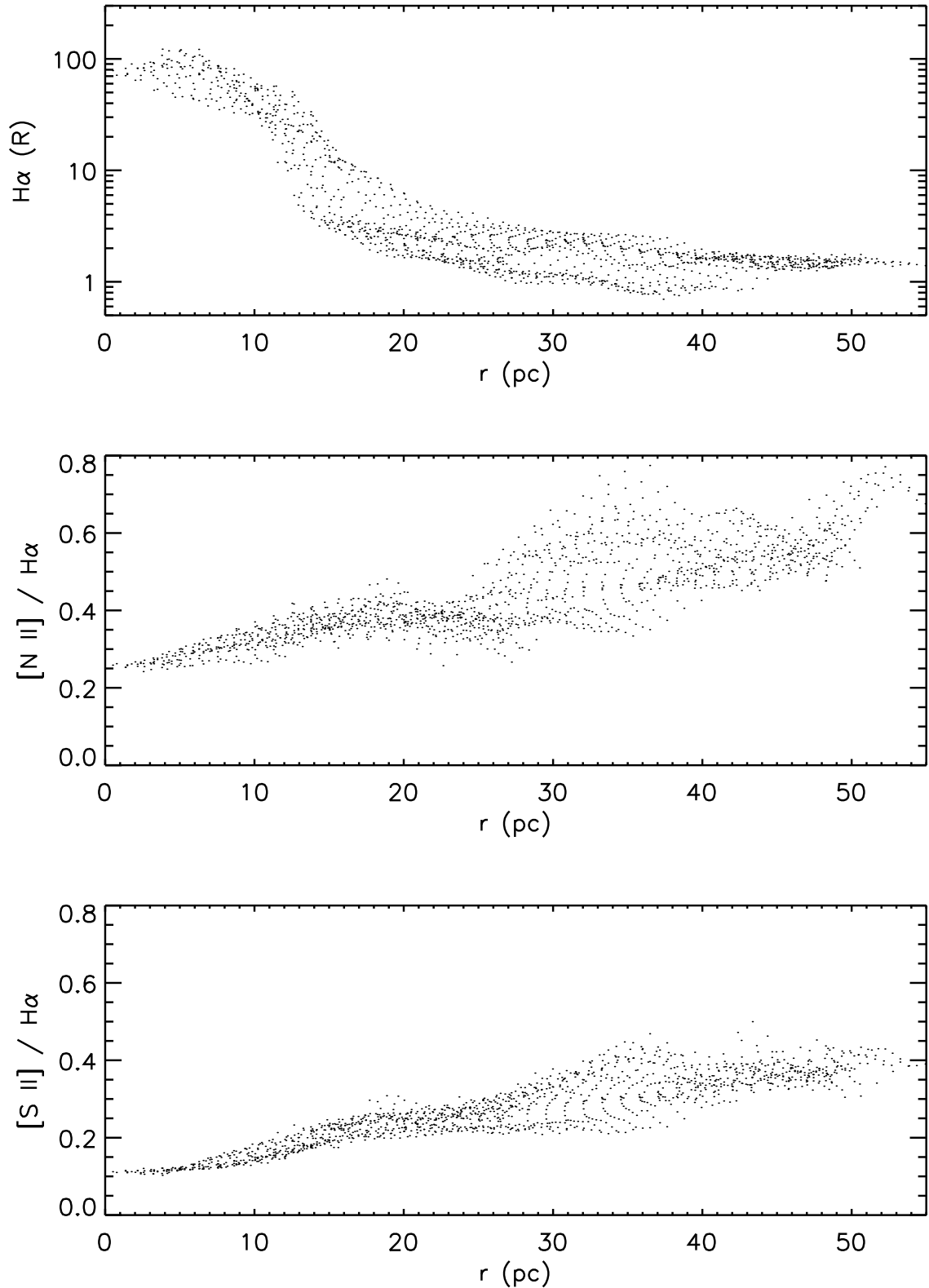}
\caption{WHAM data showing the 
variation of H$\alpha$, [N {\sc ii}]/H$\alpha$, and 
[S {\sc ii}]/H$\alpha$ as a function of radius from $\zeta$~Oph.  Notice 
the relatively smooth increase of the line ratios with distance and the 
absence of the rapid increase expected at the edge of a uniform 
density Str{\" o}mgren sphere.}
\end{figure}

\begin{figure}
\plottwo{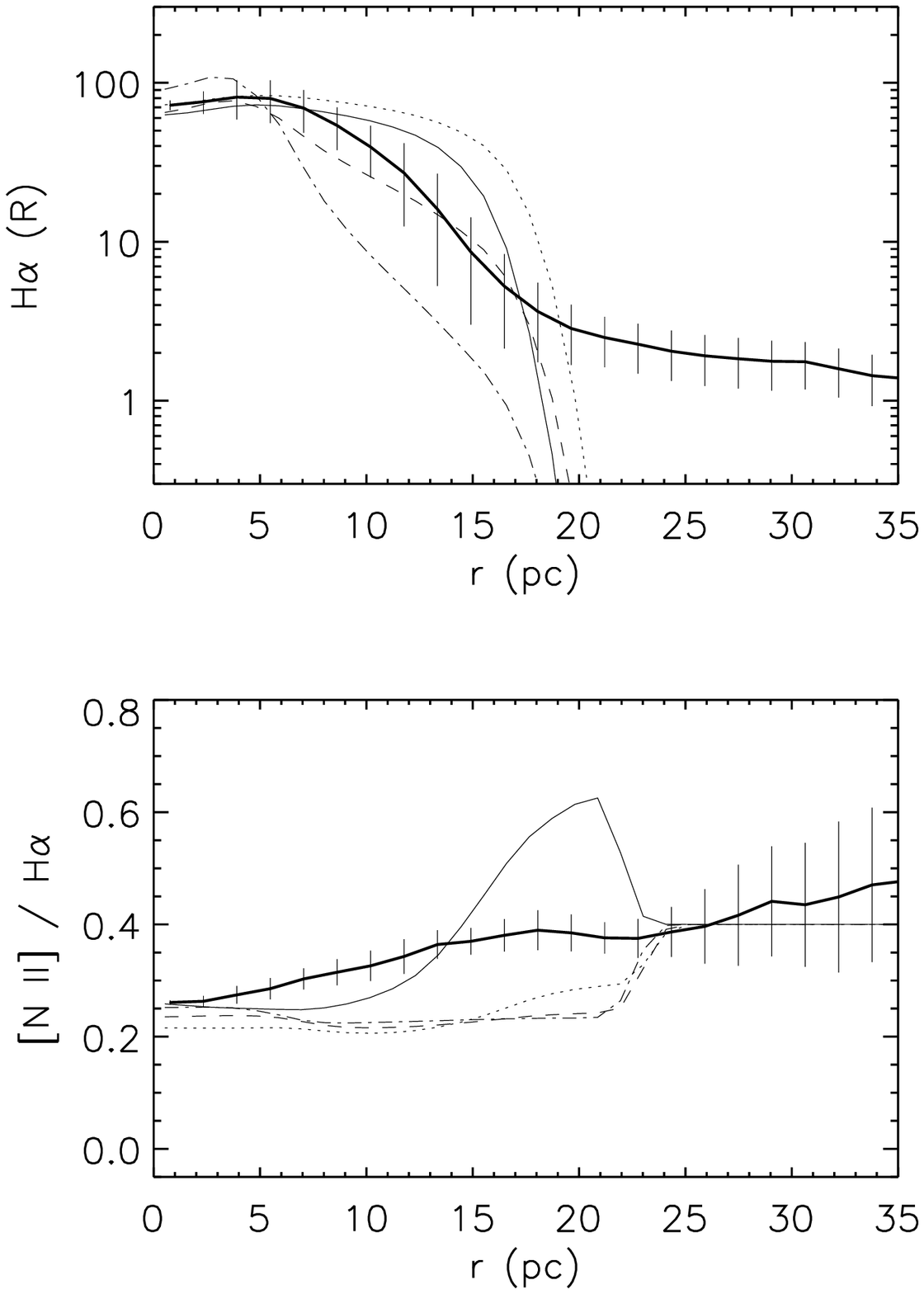}{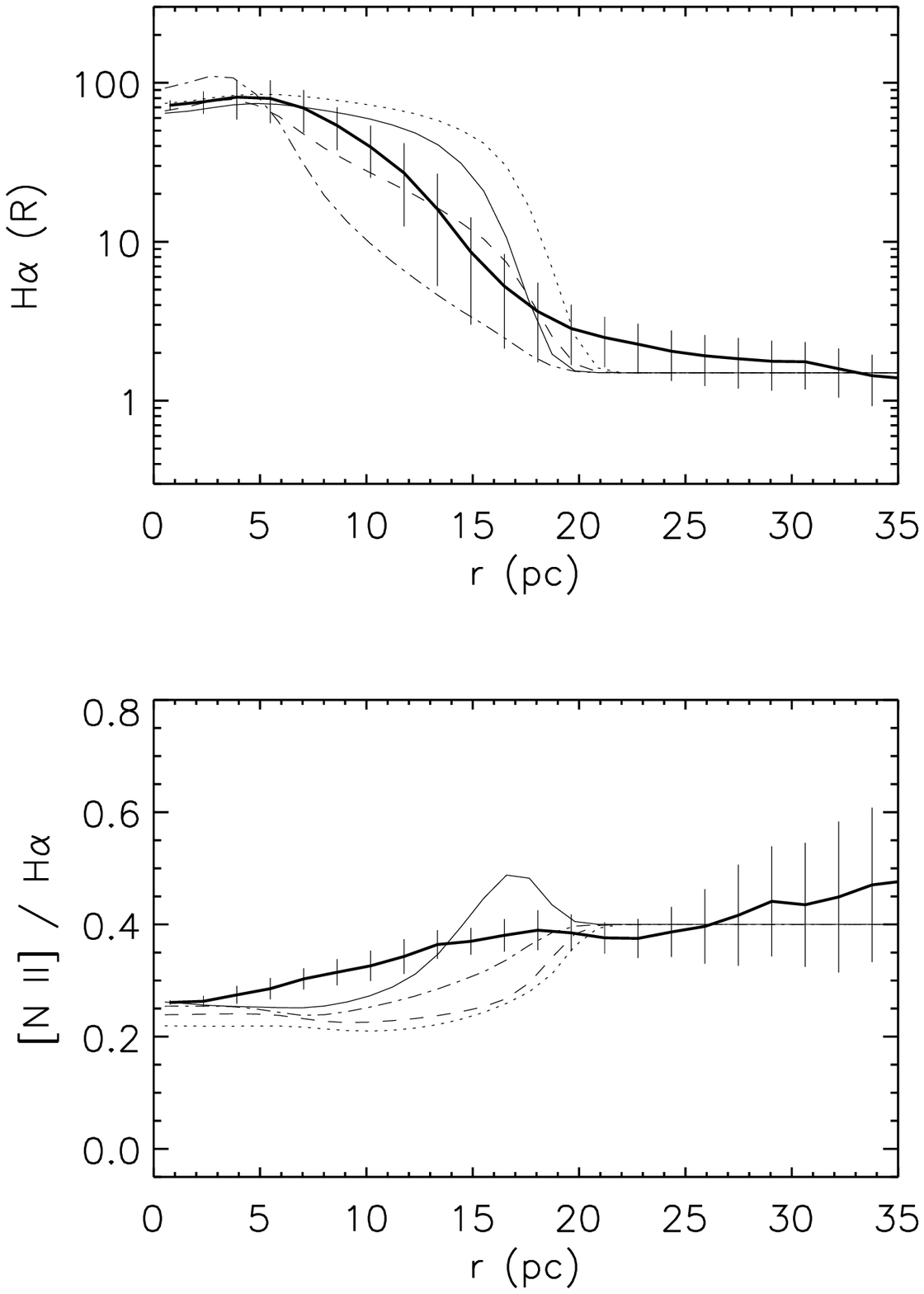}
\caption{Smooth models without (left) and with (right) the addition 
of foreground/background emission.  The heavy solid line shows the 
mean radial intensity and standard deviation (error bars) measured 
by WHAM.  The lighter lines show the models from Table~1: uniform 
ionization bounded (solid), density bounded: uniform (dotted), 
$1/r$ density (dashed), and $1/r^2$ density (dot-dashed).}
\end{figure}

\begin{figure}
\plotone{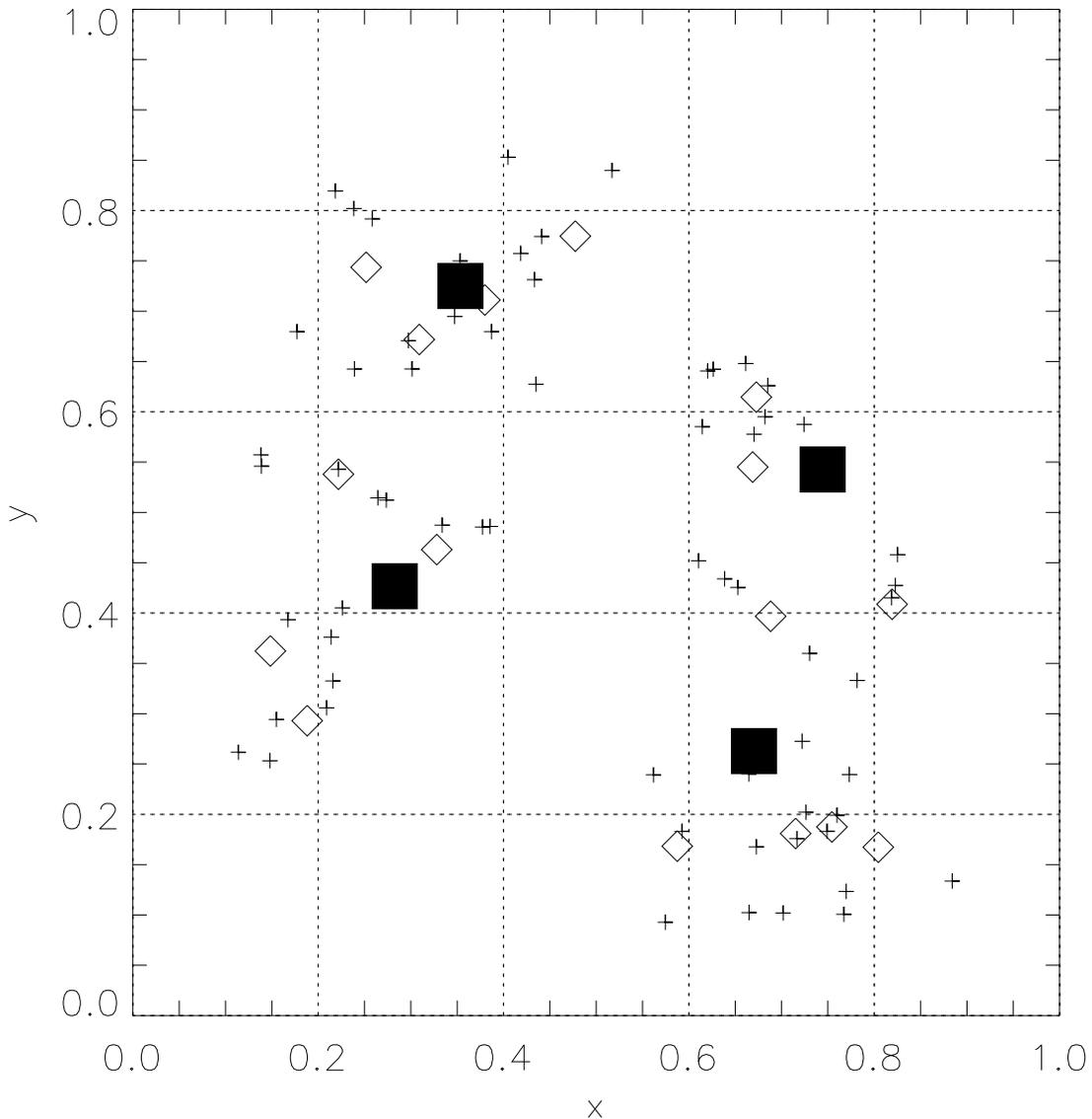}
\caption{A 2D illustration of the fractal generating algorithm for a three 
tier scheme.  Four random points (filled in squares) are cast at the first 
level.  At each subsequent level a further four points are cast around each 
of the points cast at the previous level.  In this example there are a total 
of 16 points (diamonds) cast at the second level, and a total of 64 points 
(crosses) cast at the third level.  The density in each cell in the grid 
(dashed lines) is proportional to the number of crosses in the cell.  
In our 3D models we use a five tier scheme with $N_1=32$, 64, 128, or 512 
points cast at the first level and at each subsequent level we cast 
$N=32$ points around each of the points cast at the previous level. 
Increasing $N_1$ results in a overall smoother medium (see Figs. 5 through 
8).}
\end{figure}

\begin{figure}
\plotone{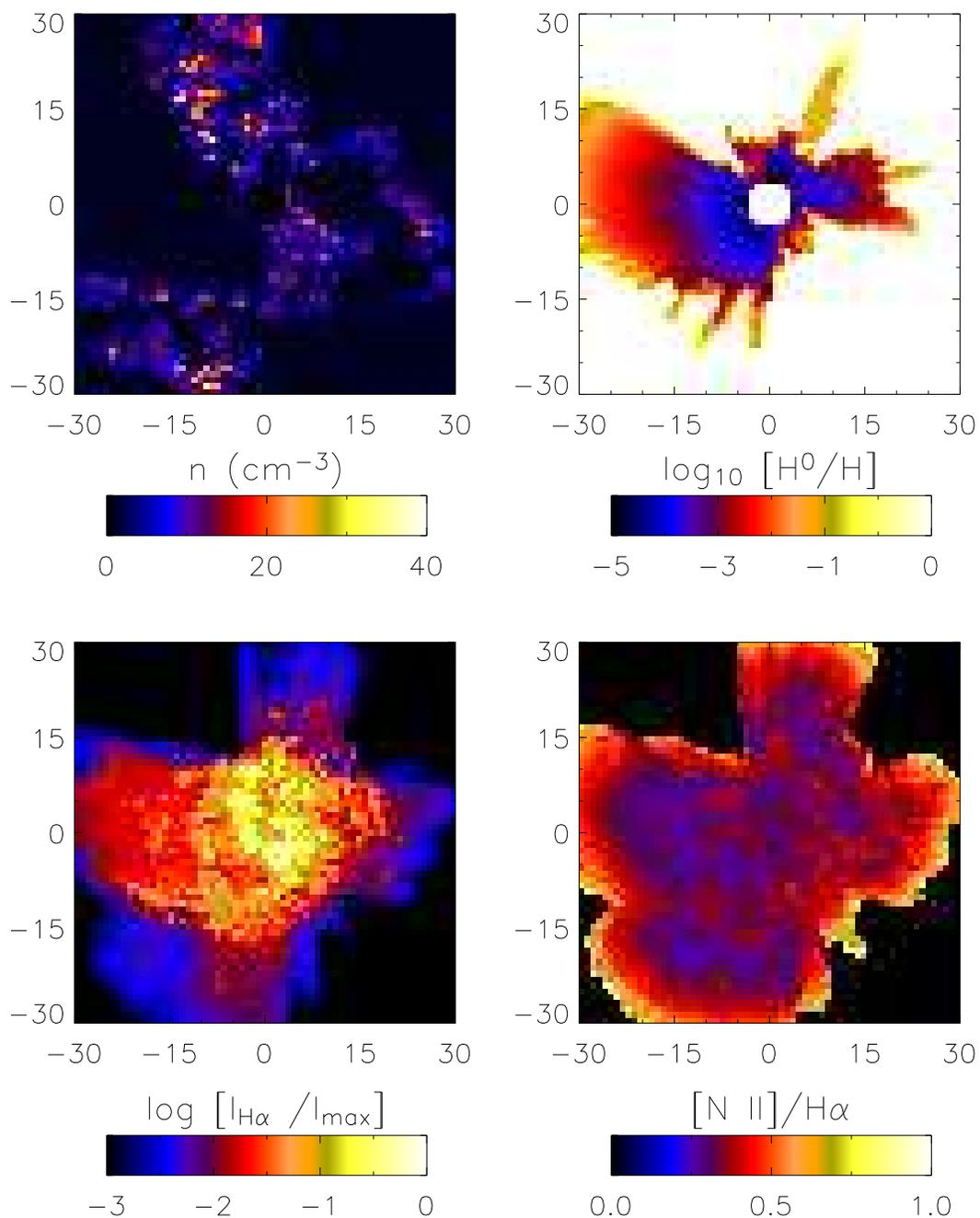}
\caption{Hierarchically clumped model with $N_1=32$ (see text) showing 
slices through the center of our grid of the number 
density and hydrogen ionization fraction (upper panels) and the 
projected H$\alpha$ intensity and projected 
[N~{\sc ii}]/H$\alpha$ line ratio maps (lower 
panels).  Axes are labeled in parsecs.  
This model allows for many sightlines from the star that do not 
intersect a dense clump, producing extended low intensity H$\alpha$ 
emission which may get lost in strong foreground/background emission.}
\end{figure}

\begin{figure}
\plotone{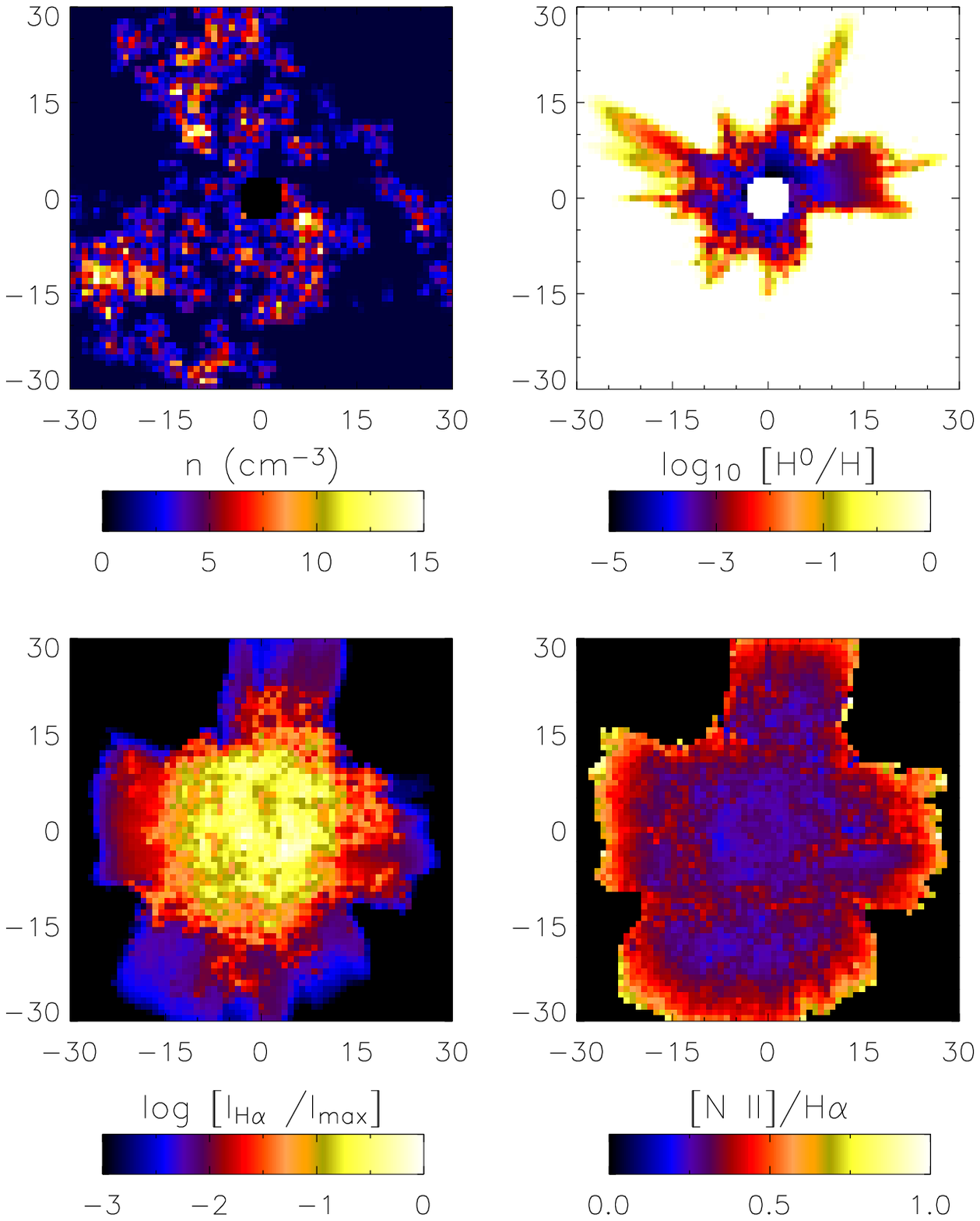}
\caption{As Fig.~4, but for $N_1=64$.}
\end{figure}

\begin{figure}
\plotone{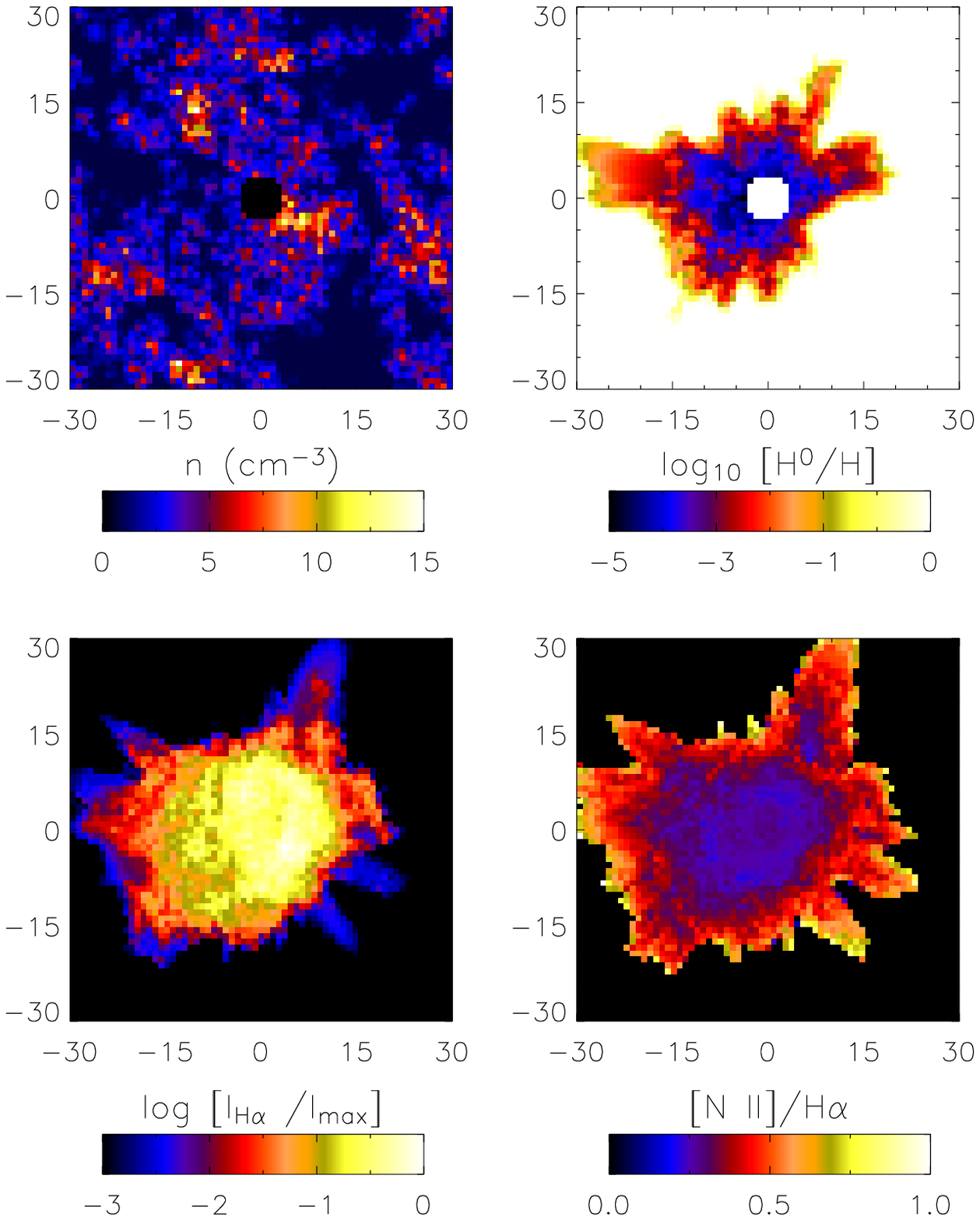}
\caption{As Fig.~4, but for $N_1=128$.}
\end{figure}

\begin{figure}
\plotone{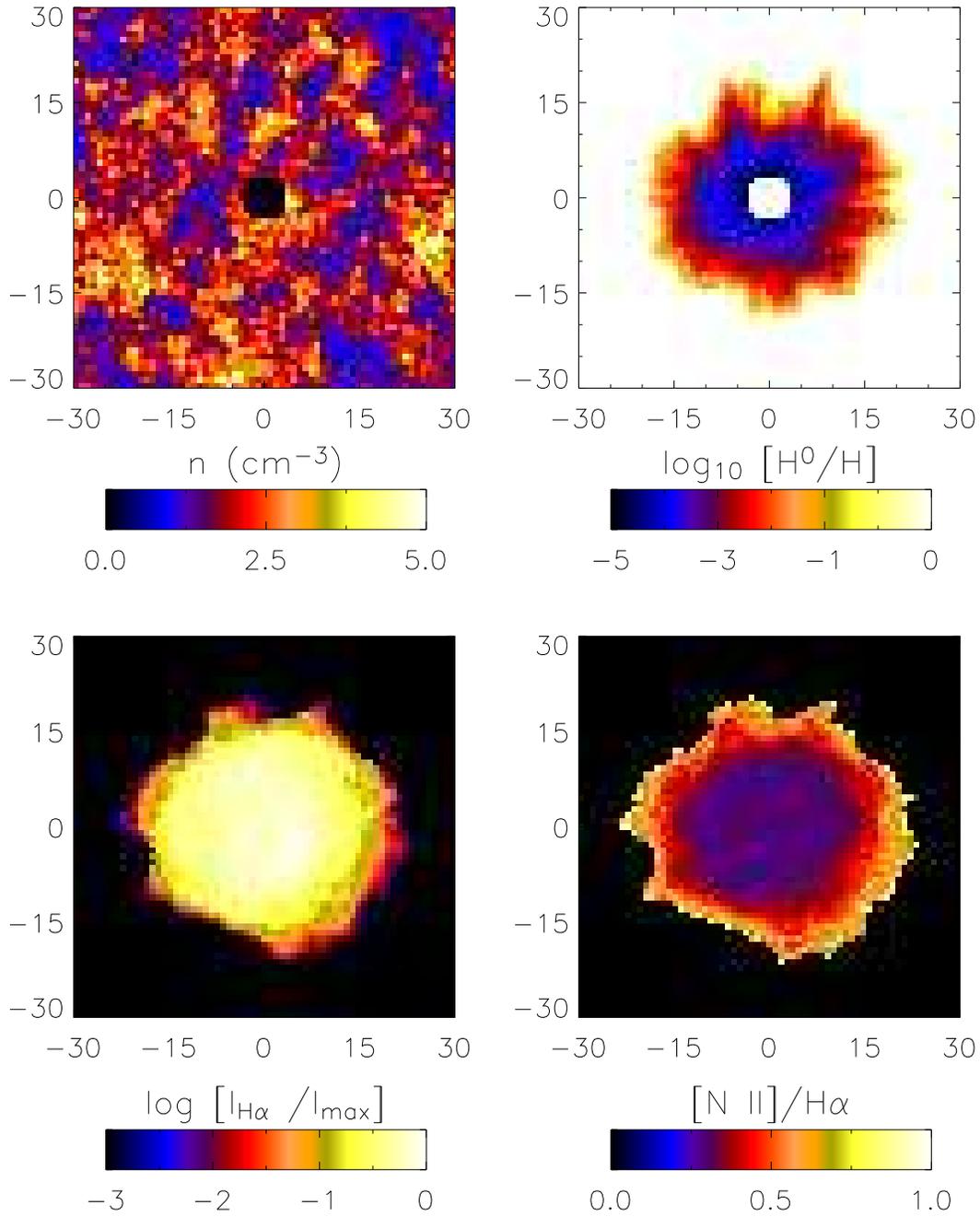}
\caption{As Fig.~4, but for $N_1=512$.  Notice how much more circularly 
symmetric the ionized region (upper right) is than in Fig.~5, because the 
many points in the initial casting tend to fill space unifromly.}
\end{figure}

\begin{figure}
\plottwo{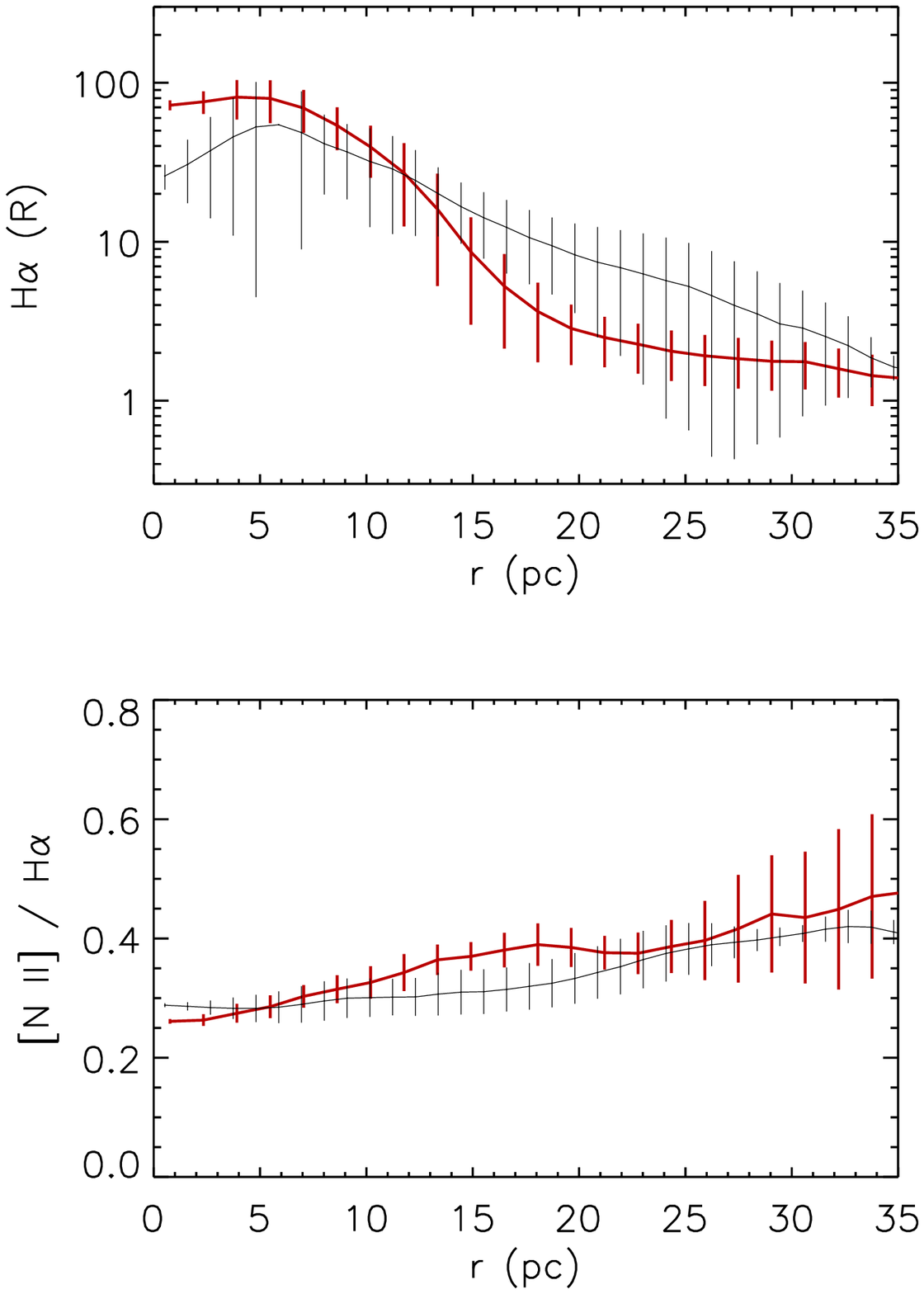}{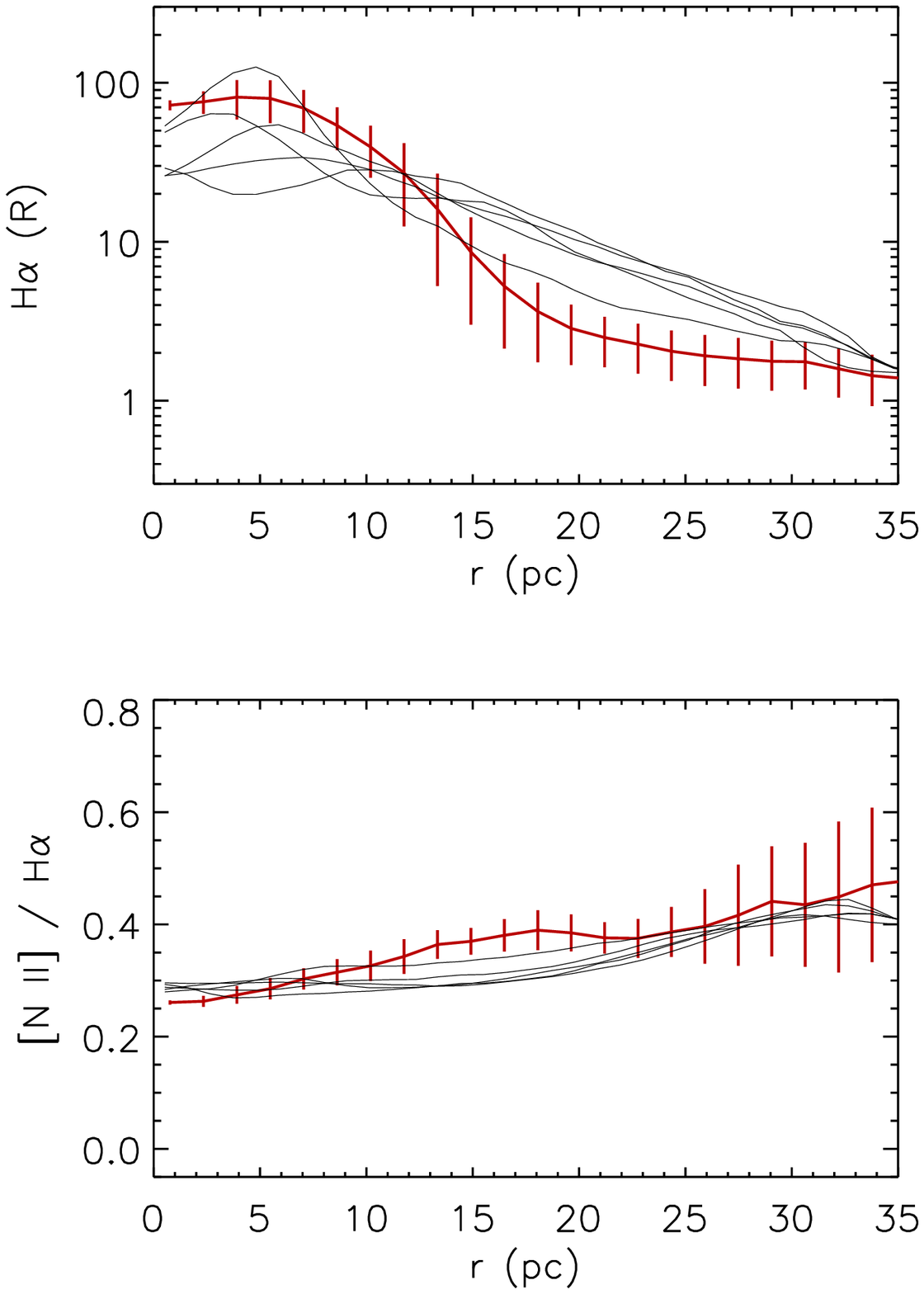}
\caption{The H$\alpha$ intensity and 
[N~{\sc ii}]/H$\alpha$ line ratios for the clumpy models with 
$N_1=32$, plotted against the projected offset distance from $\zeta$~Oph.  
Left panels: red lines 
show the mean radial intensity and standard deviations measured by WHAM 
and black lines show the same for one of the $N_1=32$ models.  A constant 
intensity of 1.5~R and 0.6~R has been added to the model H$\alpha$ and 
[N~{\sc ii}] maps to simulate the diffuse foreground/background 
emission around $\zeta$~Oph.  Right panels: red lines again show the WHAM 
data and the black lines show five different $N_1=32$ models --- each model 
has a different randomly cast hierarchical density structure.  These 
models show high H$\alpha$ at large radii due to ionizing photons 
propagating to large distances through the very porous density structure 
(see Fig.~5).  The [N~{\sc ii}]/H$\alpha$ ratio is systematically lower 
than observed and the standard deviations of the models are larger than 
in the WHAM data.}
\end{figure}

\begin{figure}
\plottwo{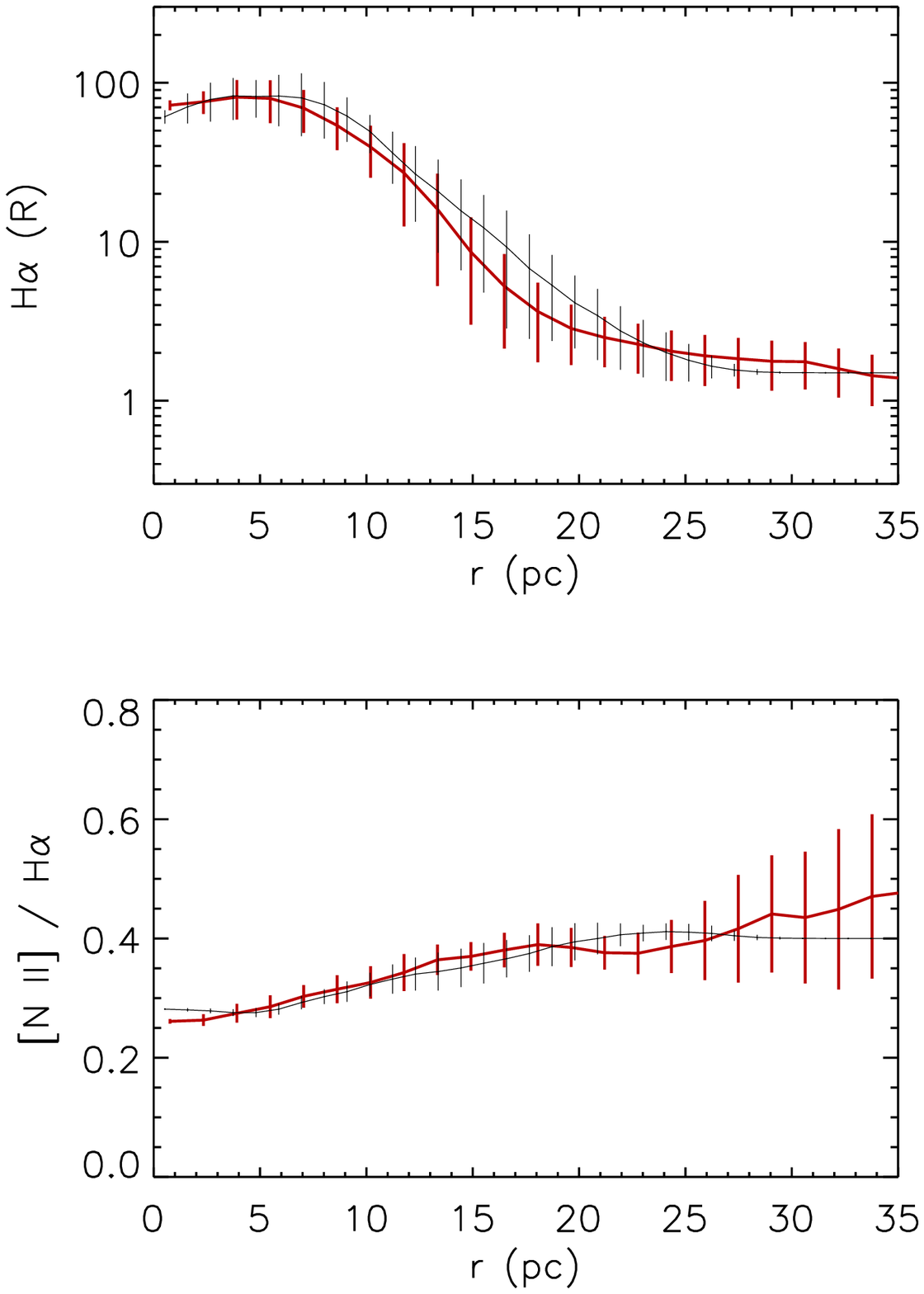}{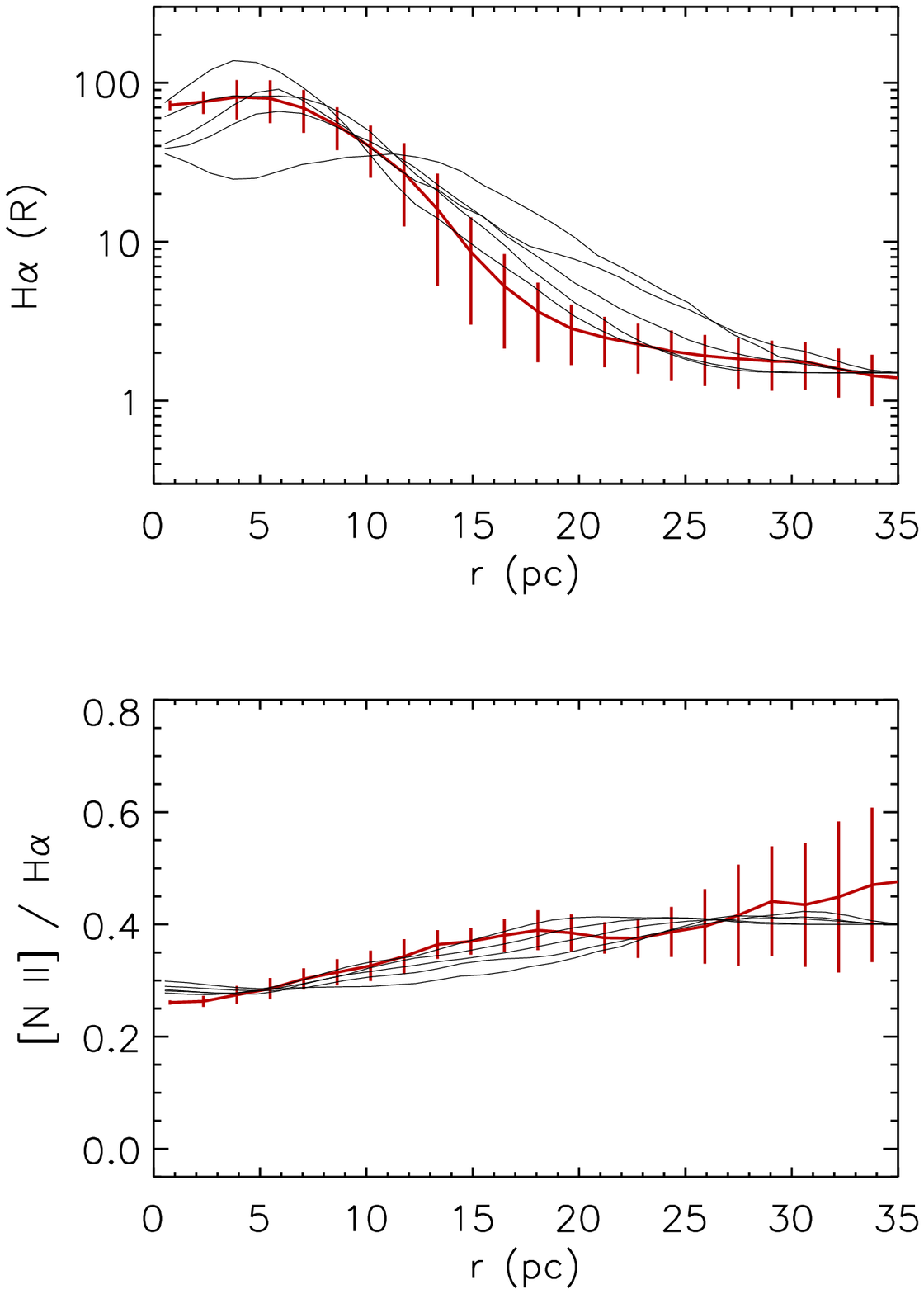}
\caption{As Fig.~8, but for $N_1=64$.}
\end{figure}

\begin{figure}
\plottwo{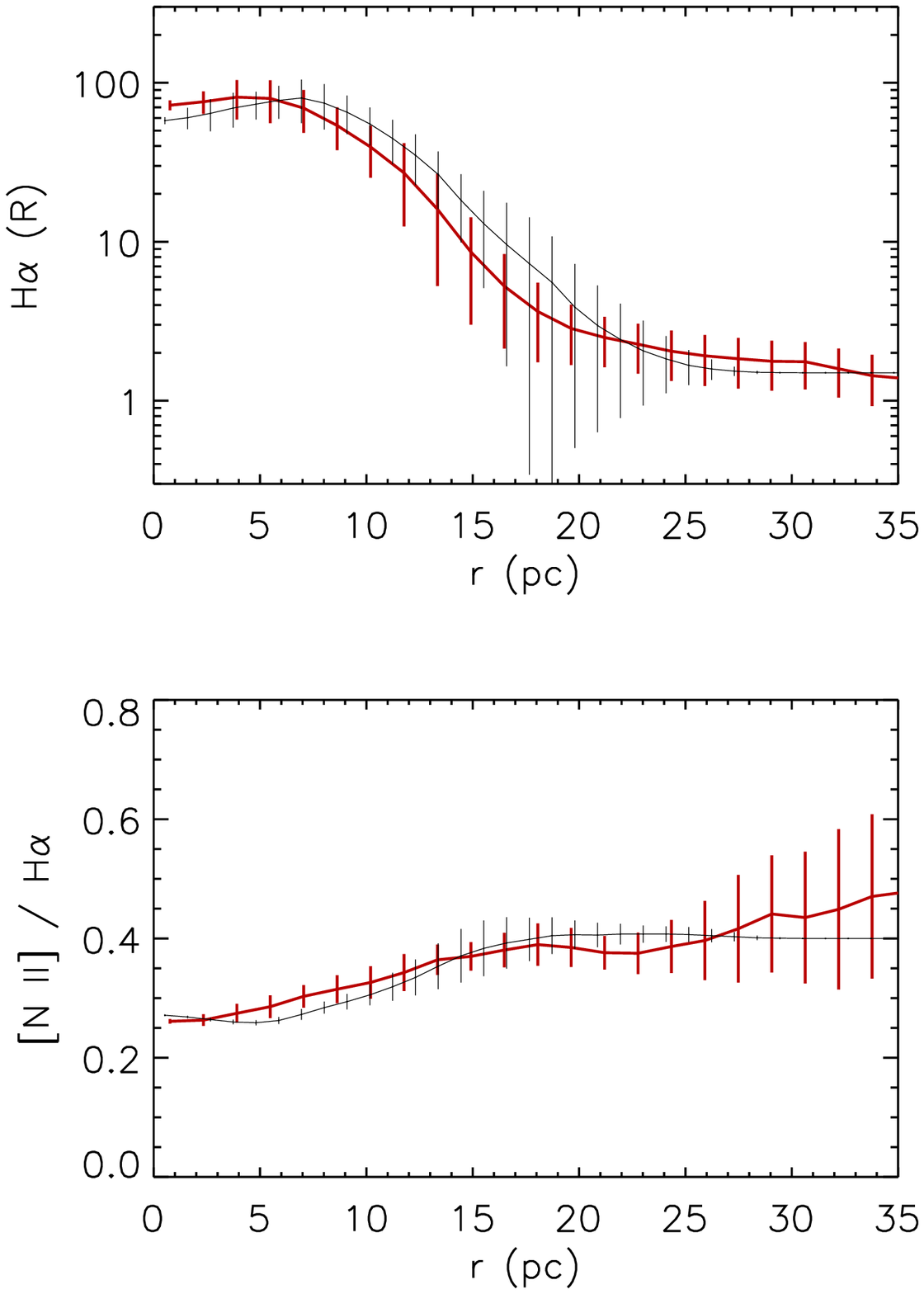}{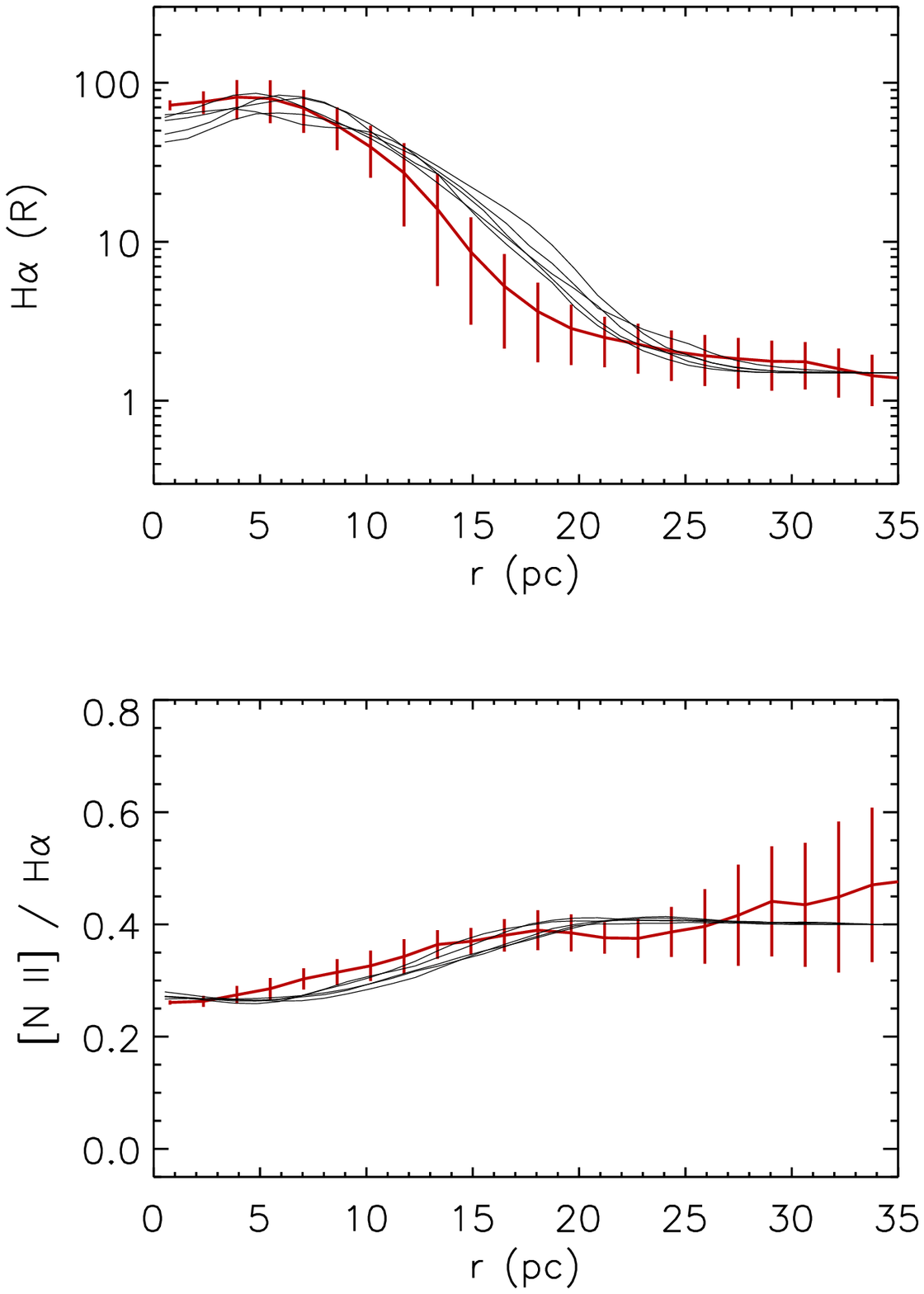}
\caption{As Fig.~8, but for $N_1=128$.}
\end{figure}

\begin{figure}
\plottwo{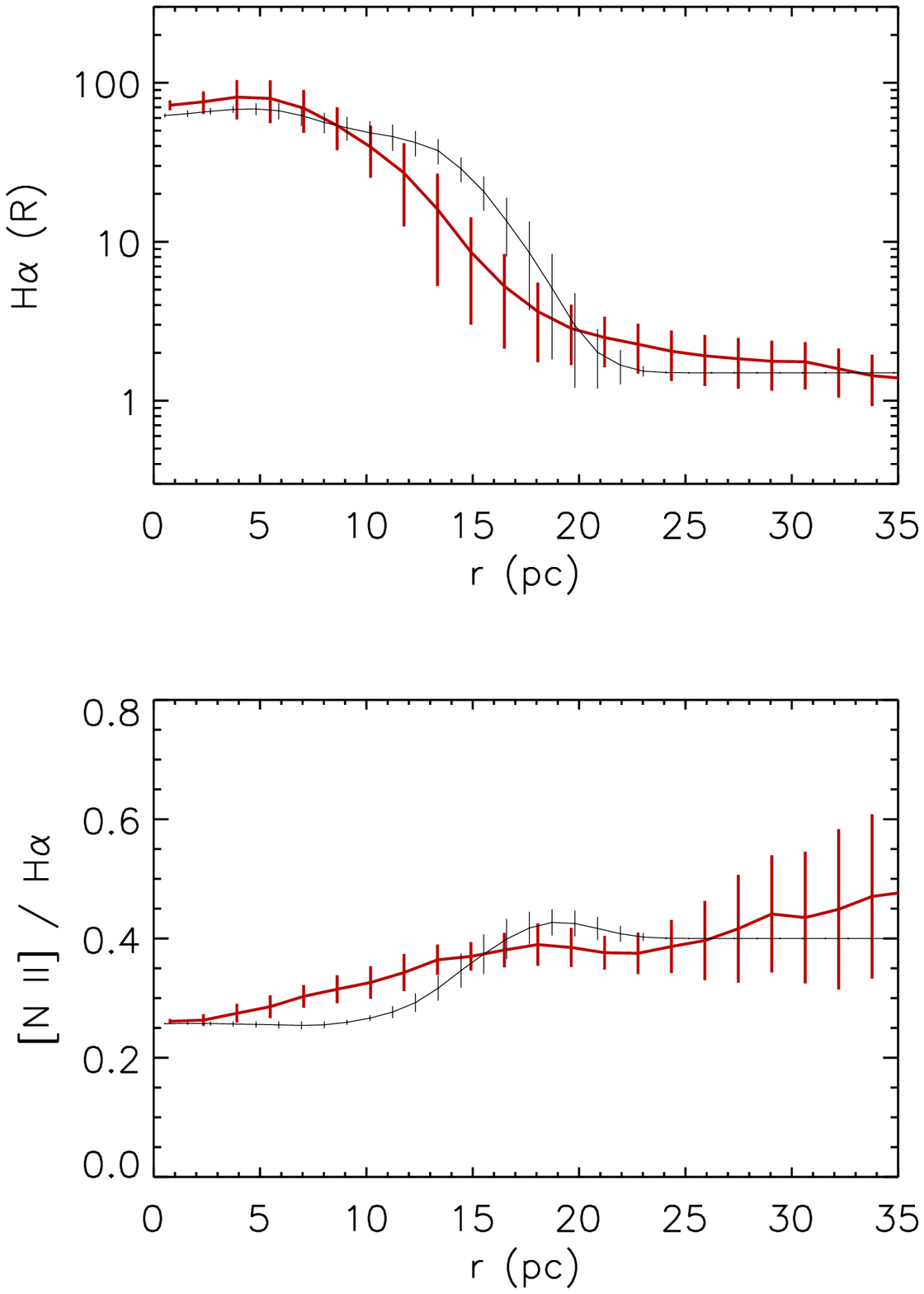}{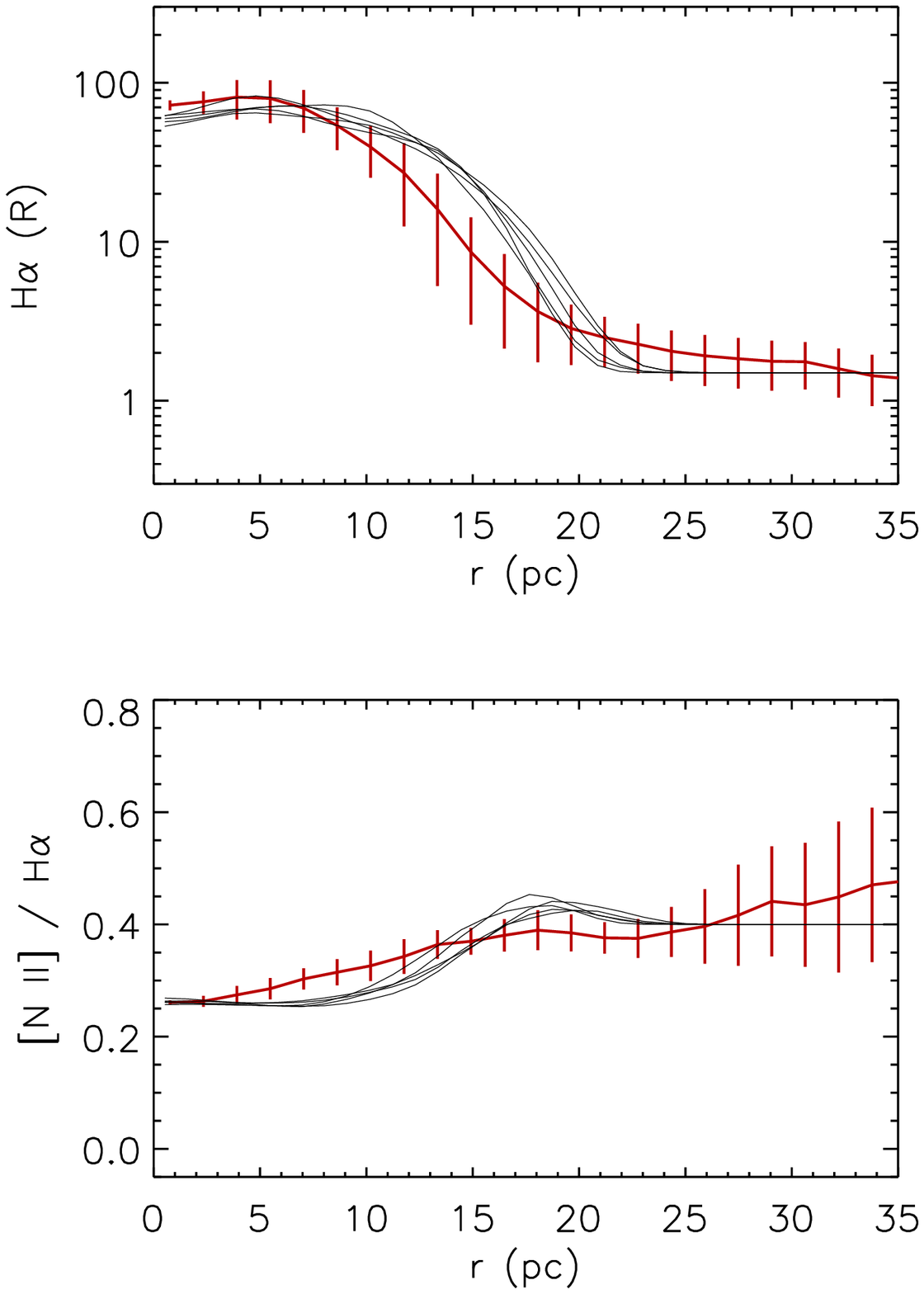}
\caption{As Fig.~8, but for $N_1=512$. 
This clumpy model closely resembles 
the uniform density, ionization bounded H~{\sc ii} region since this 
model has the smoothest density structure of the 3D models 
presented (see Figs.~5 through 8).}
\end{figure}

\end{document}